\documentclass[sigplan,screen]{acmart}

\setcopyright{cc}
\setcctype{by}
\acmDOI{10.1145/3774934.3786455}
\acmYear{2026}
\copyrightyear{2026}
\acmISBN{979-8-4007-2310-0/2026/01}
\acmConference[PPoPP '26]{Proceedings of the 31st ACM SIGPLAN Annual Symposium on Principles and Practice of Parallel Programming}{January 31 -- February 4, 2026}{Sydney, NSW, Australia}
\acmBooktitle{Proceedings of the 31st ACM SIGPLAN Annual Symposium on Principles and Practice of Parallel Programming (PPoPP '26), January 31 -- February 4, 2026, Sydney, NSW, Australia}
\received{2025-09-02}
\received[accepted]{2025-11-10}

\usepackage[figure,algo2e,linesnumbered,vlined]{algorithm2e}
\usepackage{algorithmicx}
\usepackage{algorithm}
\usepackage[noend]{algpseudocode}
\usepackage{graphicx}
\usepackage{varwidth}
\usepackage{xspace}
\usepackage{url}
\usepackage{booktabs}
\usepackage{subcaption}
\usepackage{multicol}
\usepackage{appendix}
\usepackage{placeins}
\usepackage{moresize}
\usepackage{pifont}
\usepackage{array}
\usepackage{tabularx}
\usepackage{microtype}

\newcommand{\cmark}{\textcolor{green}{\ding{52}}}%
\newcommand{\ymark}{\textcolor{yellow}{\ding{51}}}%
\newcommand{\xmark}{\textcolor{red}{\ding{56}}}%

\usepackage{xcolor}

\def\HiLi{\leavevmode\rlap{\hbox to \hsize{\color{pink!50}\leaders\hrule height .8\baselineskip depth .5ex\hfill}}}

\def\HiLiGr{\leavevmode\rlap{\hbox to \hsize{\color{lime!50}\leaders\hrule height .8\baselineskip depth .5ex\hfill}}}

\graphicspath{{./pictures/}}

\SetInd{0.2em}{0.4em}

\SetCommentSty{mycommfont}

\SetAlFnt{\scriptsize\sf}

\usepackage{mathtools}

\theoremstyle{plain}
\newtheorem{theorem}{Theorem}

\begin{document}

\title[Fixing Non-blocking Data Structures for Better Compatibility]{Fixing Non-blocking Data Structures for Better Compatibility with Memory Reclamation Schemes}

\author{Md Amit Hasan Arovi}
\orcid{0009-0009-4341-1954}
\affiliation{%
  \institution{The Pennsylvania State University}
  \city{University Park}
  \country{USA}
}
\email{arovi@psu.edu}

\author{Ruslan Nikolaev}
\orcid{0000-0002-1699-0593}
\affiliation{%
  \institution{The Pennsylvania State University}
  \city{University Park}
  \country{USA}
}
\email{rnikola@psu.edu}

\begin{abstract}
We present a new technique, Safe Concurrent Optimistic Traversals (SCOT), to address a well-known problem related to optimistic traversals with classical and more recent safe memory reclamation (SMR) schemes, such as Hazard Pointers (HP), Hazard Eras (HE), Interval-Based Reclamation (IBR), and Hyaline. Unlike Epoch-Based Reclamation (EBR), these (robust) schemes protect against stalled threads but lack support for well-known data structures with optimistic traversals, e.g., Harris' list and the Natarajan-Mittal tree. Such schemes are either incompatible with them or need changes with performance trade-offs (e.g., the Harris-Michael list).

SCOT keeps existing SMR schemes intact and retains performance benefits of original data structures. We implement and evaluate SCOT with Harris' list and the Natarajan-Mittal tree, but it is also applicable to other data structures. Furthermore, we provide a simple modification for wait-free traversals. We observe similar performance speedups (e.g., Harris vs. Harris-Michael lists) that were previously available only to EBR users. Our version of the tree also achieves very high throughput, comparable to that of EBR, which is often treated as a practical upper bound.

\end{abstract}

\ccsdesc[500]{Theory of computation~Concurrent algorithms}

\keywords{hazard pointers, non-blocking, wait-free, Harris' list, Natarajan-Mittal tree}

\maketitle

\algnewcommand{\Null}{\textbf{null}}%
\algnewcommand{\Last}{\textbf{last}}
\algnewcommand{\algorithmicgoto}{\textbf{goto}}%
\algnewcommand{\Goto}[1]{\algorithmicgoto~\ref{#1}}%
\algdef{SE}[DOWHILE]{Do}{doWhile}{\algorithmicdo}[1]{\algorithmicwhile\ #1}%
\algnewcommand\Not{\textbf{!}}
\algnewcommand\AndOp{\textbf{and}\xspace}
\algnewcommand\ModOp{\textbf{mod}\xspace}
\algnewcommand\OrOp{\textbf{or}\xspace}

\SetKw{Break}{break}
\SetKw{Continue}{continue}
\SetKwIF{If}{ElseIf}{Else}{if (}{)}{else if (}{else}{endif}

\SetKwRepeat{Do}{do}{while}
\SetKwProg{Fn}{}{}{}
\newcommand{\removelatexerror}{\let\@latex@error\@gobble}

\algnewcommand{\LineComment}[1]{\State \(\triangleright\) #1}

\SetKwRepeat{Struct}{struct \{}{\}}
\SetKwRepeat{TemplateStruct}{template <T> struct \{}{\}}

\section{Introduction}
With the rise of multicore and manycore systems, parallelization becomes essential for improving performance across the entire software stack, from applications to operating system (OS) components.
Non-blocking data structures, which are not dependent on mutual exclusion, are becoming increasingly popular due to their scalability, throughput, and latency characteristics. Unfortunately, many data structures that utilize non-blocking techniques conflict with fundamental assumptions of classical memory management. This is because memory cannot be promptly reclaimed after an object is removed from a data structure, as there may still exist stale pointers to the object due to ongoing operations.
When using manual memory management, memory has to be safely reclaimed by freeing memory only when \emph{all} ongoing operations with stale pointers complete. A number of \emph{safe memory reclamation} (SMR) techniques were proposed for C, C++, and Rust.
While we do not want to compromise data structure progress properties, it can be challenging to achieve several \emph{highly desirable} properties simultaneously with non-blocking progress in SMR.
A recent work~\cite{ERATHEOREM} sheds more light onto this conundrum: it proves that it is \emph{impossible} to achieve three highly desirable SMR properties simultaneously, i.e., only two properties can ever be achieved: \textbf{(A) robustness} (bounded memory usage), \textbf{(B) easy integration} (no dependency on any special APIs, such as roll-back mechanisms or OS-based primitives), and \textbf{(C) wide applicability} (compatibility with most data structures without changes).

While striving to have (A) for non-blocking progress,\footnote{If memory usage is unbounded, no further system progress can be made once memory is exhausted, regardless of whether actual locks are present.} many researchers also prioritize (C) because, intuitively, their goal is to support as many data structures as possible.
However, we argue that sacrificing (C) rather than (B) by carefully adapting existing problematic data structures is a ``necessary evil,'' which is a \emph{crucial new insight}. First, (C) in its strongest form -- applying to \emph{all} data structures -- is only currently known to exist in \emph{blocking} SMR schemes~\cite{ERATHEOREM}.
Second, even if we accept (C) in its relaxed form -- applying to \emph{most} data structures -- the absence of (B) would lead to either inability to achieve strict non-blocking progress in practice due to OS dependency~\cite{debra,NBR,Balmau,forkScan,threadScan}, lack of universally-defined APIs~\cite{VBR}, or both.
Third, the lack of (C) implies that adaptations are required, not impossibility of the implementation.

EBR~\cite{epoch1} is a fast and easy-to-use approach but has a serious drawback that any stalled thread results in unbounded memory usage, i.e., it lacks (A). Once exhausting memory, the system must either reboot or block until memory is freed. HP~\cite{hazardPointers} is an antipode of EBR: it is slower and more difficult to use but provides robustness with strict non-blocking guarantees.
Prior to this work, well-known data structures such as Harris' linked list~\cite{HarrisList} with optimistic traversals and the Natarajan-Mittal binary search tree (BST)~\cite{NMTree} were not known to work well with HP~\cite{HPPP,DRC},\footnote{A few issues were discussed in~\cite{DRC}, but only the memory leak was addressed.} which is due to HP's lack of the wide-applicability (C) property. A number of other recent schemes, including Interval Based Reclamation (IBR)~\cite{IBR}, Hyaline-1S~\cite{HYALINEPLDI}, and Hazard Eras (HE)~\cite{HEBenchmark}, sometimes offer better performance but have similar issues to HP, though IBR and Hyaline-1S are generally easier to use in practice. 
HP++~\cite{HPPP} is another recent scheme. HP++ is slower than HP and, strictly speaking, makes compromises with both (B) and (C), but it still supports these two data structures.
We are inspired by HP++'s success, which indicates that \textbf{many data structures can still be implemented}, despite earlier views about the limited applicability of SMR schemes in this category.

This paper introduces an alternative method, Safe Concurrent Optimistic Traversals (SCOT). We depart from the typical strategy of designing a ``silver-bullet'' SMR approach for the reasons described above. Instead, we argue that the problematic data structures can be modified to accommodate SMR schemes with strict non-blocking properties.
SCOT's \emph{key insight} is that optimistic traversals are still feasible with HP and other SMR schemes as long as at every traversal step, we perform a simple safety check, which allows us to proceed safely to the next data-structure object.
We also discuss how to address our approach's limits for \emph{wait-free} traversals.

We present SCOT for Harris' list and the Natarajan-Mittal tree and evaluate these data structures with HP, HE, IBR and Hyaline-1S. We also compare results to EBR and show that performance benefits can be preserved in the same manner.

\section{Background}

\label{sec:background}

\subsection{Non-Blocking Progress Guarantees}
In the literature, several \emph{non-blocking} progress properties are considered. With \emph{obstruction-freedom}, progress is only guaranteed when a thread runs in isolation
from others. \emph{Lock-freedom} allows threads to interfere with each other while still guaranteeing that \emph{at least} one thread makes progress in a finite number of steps. (It should be noted that good lock-free algorithms allow multiple threads to make concurrent progress in practice.)
This should not be confused with ``lockless'' approaches that simply avoid \emph{explicit} locks, where one preempted thread can still block all other threads.
\emph{Wait-freedom} is a much stronger property which requires that \emph{all} threads make progress in a finite number of steps.

Most non-blocking data structures use \emph{compare-and-swap} (CAS) for synchronization. CAS is an instruction which takes three arguments: a \emph{pointer} to the shared variable to be updated, an \emph{expected} value, and a \emph{desired} new value; CAS atomically replaces the shared-variable value with the desired value if its current value equals the expected value.

\subsection{Memory Reclamation}
All SMR schemes used in this paper follow the same lifecycle: a removed node is first \emph{retired}, and then reclaimed only once the scheme guarantees that no thread can access it anymore. Each SMR scheme maintains per-thread metadata (epochs, eras, or reserved pointers) to determine reclamation safety. Although SMR mechanisms differ, they implement the same two-step process: unlink and retire, then reclaim when safe.

\subsubsection{Epoch-Based Reclamation (EBR)}
Epoch-Based Reclamation (EBR)~\cite{epoch1, epoch2} ensures memory safety using a global epoch. Threads entering a critical section publish the current epoch, and retired nodes are stored in per-thread lists tagged with that epoch. The global epoch advances periodically, and nodes are reclaimed only after all active threads have moved beyond the retire epoch. The main limitation of EBR is its sensitivity to stalled or crashed threads, since the non-progressing thread can prevent epoch advancement and lead to unbounded memory accumulation.

\subsubsection{Hazard Pointers (HP)}\label{sec:backgroundhp}
Hazard Pointers (HP)~\cite{hazardPointers} ensure memory safety by requiring each thread to publish any pointer it may dereference in a globally visible hazard slot. A retired object is reclaimed only after scanning all slots to verify that no thread still references it. This makes HP robust to stalled or preempted threads because it does not rely on global quiescence. The approach introduces overhead due to memory barriers and hazard-slot scans, and it also complicates traversal-heavy code. Despite these costs, HP remains widely used for strong safety guarantees.

In Section~\ref{sec:design}, we rely on the \texttt{protect} and \texttt{dup} operations (shown in Figure~\ref{alg:hpprotect}).
\verb|protect| safely retrieves an object by making a \emph{reservation} in the global array; the idea is that other threads will not reclaim reserved pointers. Each local pointer has a unique \emph{index}. Due to a small race window, HP verifies in a loop that the pointer has not changed after the reservation was made.
\texttt{dup} copies an already protected pointer from one hazard slot to another, enabling safe transitions between traversal roles (e.g., from the \emph{next} to the \emph{current} pointer).
This is crucial to avoid transient unprotected states.
(An alternative to \verb|dup| is index renaming via an indirection array, but from our observations, \verb|dup| is generally cheaper.)

\begin{figure}
\begin{minipage}{.57\columnwidth}
\begin{algorithm2e}[H]
\Fn{\upshape \textbf{node\_t *}protect(\textbf{node\_t*} \&var, \textbf{int} idx)}
{
    \textbf{node\_t *}ret, \textbf{*}n = \textbf{nullptr}\;
    \While{\upshape ( (ret = var.\textbf{load}()) != n)}{
        hp[TID][idx] = ret\tcp*{Also clear}
        n = ret\tcp*{logical-deletion bits}
    }
    \Return ret\;
}
\end{algorithm2e}
\end{minipage}
\begin{minipage}{.42\columnwidth}
\begin{algorithm2e}[H]
\setcounter{AlgoLine}{6}
\tcp{Duplicate $HP_i$ to $HP_j$}
\Fn{\upshape \textbf{node\_t *}dup(\textbf{int} i, \textbf{int} j)}
{
    \textbf{node\_t*} p = hp[TID][i]\;
    \tcp{Release semantics}
    hp[TID][j] = p\;
    \Return p\;
}
\end{algorithm2e}
\end{minipage}
\caption{HP methods. (TID is the current thread number.)}
\label{alg:hpprotect}
\end{figure}

\subsubsection{Hazard Eras (HE)}
Hazard Eras (HE)~\cite{HEBenchmark} optimize hazard pointers by using logical timestamps (``eras'') instead of explicit pointer protection. Threads record the current era when accessing an object, and retired objects are tagged with a retire era; reclamation is safe once no thread holds an access era that intersects with it. This temporal mechanism is an alternative to the direct pointer mechanism and reduces the number of memory barriers compared to HP. 

\begin{table}[htbp]
\caption{Subjective performance characterization of common non-blocking data structures and their compatibility with SMRs. (* HP column also applies to HE/IBR/Hyaline-1S.)}
\begin{center}
\resizebox{\columnwidth}{!}{
\begin{tabular}{l c c c c}
    \toprule
    \textbf{Data Structure} & \textbf{Fast} & \textbf{EBR} & \textbf{HP*} & \textbf{HP* with SCOT}  \\
    \midrule
    Harris' Linked List~\cite{HarrisList} & \cmark & \cmark & \xmark & \cmark \\
    Harris-Michael Linked List~\cite{10.1145/564870.564881} & \ymark & \cmark & \cmark  & \cmark \\
    Fraser's Skip List~\cite{epoch1} & \cmark & \cmark & \xmark & \cmark \\
    Herlihy-Shavit Skip List~\cite{Herlihy:2012:AMP} & \ymark & \cmark  & \cmark & \cmark \\
    Natarajan-Mittal Tree~\cite{NMTree} & \cmark & \cmark & \xmark & \cmark \\
    Ellen et al. Tree~\cite{ELLENTREE} & \xmark & \cmark & \cmark & \cmark \\
    \bottomrule
\end{tabular}
}
\end{center}
\label{tbl:compareds}
\end{table}

\subsubsection{Interval-Based Reclamation (IBR)}
Interval-Based Reclamation (IBR)~\cite{IBR} extends era-based methods by tracking access \emph{intervals} rather than individual accesses. Each thread maintains a birth and retire era, and a retired node is reclaimable once no thread's interval overlaps with its lifetime. IBR avoids explicit tracking of local pointers via indices, which simplifies its programming model. 

\subsubsection{Hyaline-1S}
Hyaline-1S~\cite{HYALINEPLDI} is another SMR technique, which adopts the birth era approach from HE and IBR. Instead of retire eras, it maintains an object (reference) counter.  Hyaline-1S reduces memory barriers by using reference counting only during reclamation. Each thread maintains a birth era, and an object is reclaimed only when its implicit lifetime interval does not overlap with any active thread's era, ensuring safety even with stalled threads. Reclamation is done by \emph{any} thread, improving the performance. 

\subsection{Overview of Data Structures}
Table~\ref{tbl:compareds} characterizes common data structures by speed and SMR compatibility, both original and with SCOT. We omit hash maps as they are simply arrays of Harris' or Harris-Michael lists.
These data structures employ ``logical'' deletion, i.e., a node inside a data structure is first simply marked (by stealing one bit from a node pointer), and then unlinked from the data structure (i.e., ``physically'' deleted). Harris' list and Fraser's skip list can use read-only\footnote{The original version~\cite{HarrisList} lacks the read-only optimization from~\cite{Herlihy:2012:AMP}.} optimistic traversals, which allow for safely continuing the iteration through the data structure, even when encountering logically deleted nodes. Moreover, a chain of \emph{consecutive} logically deleted nodes is retired using one CAS operation. These optimizations improve performance and reduce contention.

However, as we discuss below, this approach is \emph{incompatible} with HP. Harris-Michael lists and Herlihy-Shavit skip lists~\cite{Herlihy:2012:AMP} use a different approach: they still use ``logical'' deletion. However, they mandate that logically deleted nodes are removed \emph{immediately} after the first encounter and before proceeding further, including the search operation.
Ellen et al. tree and the Natarajan-Mittal tree operate on tree branches rather than individual nodes. However, there is a similar distinction between them with respect to optimistic traversals and ability to remove multiple ``logically'' deleted branches.

\begin{figure}
    \centering
    \includegraphics[width=\columnwidth]{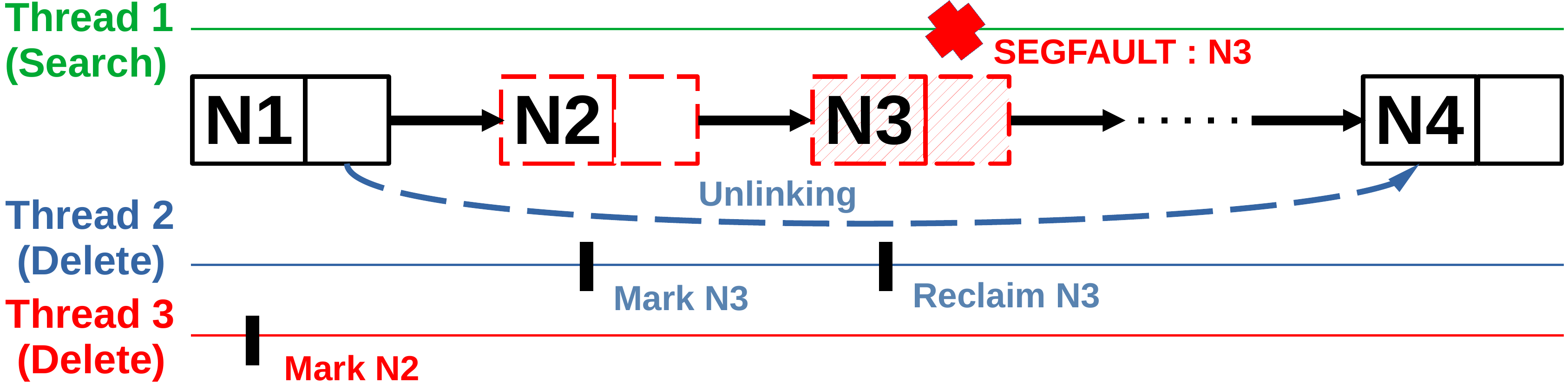}
    \caption{Unsafe optimistic traversal of Harris' linked list with HP: accessing nodes after N2 may cause SEGFAULTs.}
    \label{fig:SCOT_HL_HP}
\end{figure}

As we show in the table, data structures with optimistic traversals are incompatible with HP, HE, IBR, and Hyaline-1S, a problem that we address in this paper.

\subsection{Harris' and Harris-Michael Linked Lists}
The main challenge in lock-free linked lists is allowing a deleting thread to mark a node as deleted and update the preceding node's next pointer.
Timothy L. Harris solved this~\cite{HarrisList} by using helping threads during deletion. A deleting thread first marks a node as ``logically'' deleted by updating its next pointer, and then the node is ``physically'' removed by the deleting thread or a helper. (\emph{Search} jumps over logically deleted nodes, so it is not a conflicting operation, but a concurrent \emph{Insert} or \emph{Delete} that requires the physical removal of the node is).
Harris' list supports \emph{optimistic traversals}, meaning that logically deleted nodes can still be traversed and bypassed by a search operation without immediate physical removal. This allows a thread to defer unlinking of logically deleted nodes until it reaches the intended destination node.

However, this approach poses challenges for HP (Figure~\ref{fig:SCOT_HL_HP}). Suppose Thread 3 marks N2 for deletion but has not yet unlinked it; Thread 2 marks N3 and unlinks the entire chain between N1 and N4 (assuming all \emph{consecutive} nodes are logically deleted). Meanwhile, Thread~1, traversing the list, reaches N2 before Thread 2 physically removes the chain and retires its nodes. Although Thread~2 retires N2, it remains in the HP \emph{limbo list} (i.e., the list of retired but not-yet reclaimed nodes) since Thread 1 reserved it \emph{before the physical deletion}. Thread 2 retires N3, but since Thread 1 never reserves it, N3 is reclaimed causing Thread 1 to fail when accessing it.

Maged M. Michael modified Harris' approach~\cite{10.1145/564870.564881} to address its incompatibility with HP: when a thread encounters a logically deleted node during a traversal, it \emph{immediately} attempts to physically remove the node.

\textbf{Why Michael's Approach Works.} Accessing N2 is safe since its HP reservation precedes physical deletion. (To \emph{successfully} make a reservation, HP verifies that the pointer from N1 to N2 remains intact.) After N2 is physically deleted, it is retired and placed into the HP limbo list.
If we now attempt to reserve N3, HP erroneously succeeds as \emph{the pointer from N2 to N3 stays intact, unlike the pointer from N1 to N2}. Thus, Michael's approach ensures that the successor of a marked node is never traversed: the marked node is unlinked from the list and the operation is restarted if the unlinking fails.\footnote{Another way to think about it is that if we remove nodes one by one, N3 takes place of N2, i.e., right after N1. N3 will be the first logically deleted node in the chain when making HP reservation, which is still safe to access.}

While solving the HP problem, Michael's method increases the number of CAS operations, which increases thread contention and makes read-only wait-free traversals impossible.

\begin{figure}
\begin{minipage}{.46\columnwidth}
\begin{algorithm2e}[H]
\SetAlgoNoLine
\Struct{\upshape node\_t} {
    \textbf{node\_t *} Next\tcp*{Next node}
    \textbf{key\_t} Key\tcp*{Any key type}
}
\textbf{node\_t} * Head\tcp*{List head}
\SetAlgoVlined
\Fn{\upshape \textbf{bool} Insert(\textbf{key\_t} key)} {
  \textbf{node\_t} *new = malloc(\textbf{sizeof}(node\_t))\;
  new->Key = key\;
  \textbf{node\_t} **prev, *curr, *next\;
  \While {\upshape \textbf{true}} {
    \If {\upshape Do\_Find(key, \&prev, \&curr, \&next, \textbf{false})} {
       free(new)\;
       \Return \textbf{false}\;
    }
    new->Next = curr\;
    \lIf {\upshape CAS(prev, curr, new)} {
       \Return \textbf{true}}
}
}
\Fn{\upshape \textbf{bool} Delete(\textbf{key\_t} key)} {
  \textbf{node\_t} **prev, *curr, *next\;
  \While {\upshape \textbf{true}} {
    \If {\upshape !Do\_Find(key, \&prev, \&curr, \&next, \textbf{false})} {
       \Return \textbf{false}\;
    }
    \lIf {\upshape !CAS(\&curr->Next, next, getMarked(next))} {
       \textbf{continue}}
    \lIf* {\upshape CAS(prev, curr, next)} {}
    \tcp{Delete curr}
    \Return \textbf{true}\;
    }
}
\end{algorithm2e}
\end{minipage}
\begin{minipage}{.53\columnwidth}
\begin{algorithm2e}[H]
\setcounter{AlgoLine}{23}
\Fn{\upshape \textbf{void} Init()} {
  Head = malloc(\textbf{sizeof}(node\_t))\;
  Head->Next = \textbf{nullptr}\;
  Head->Key = $\infty$\;
}
\Fn{\upshape \textbf{bool} Search(\textbf{key\_t} key)} {
  \textbf{node\_t **}prev, \textbf{*}curr, \textbf{*}next\;
  \Return Do\_Find(key, \&prev, \&curr, \&next, \textbf{true})\;
}
\Fn{\upshape \textbf{bool} Do\_Find(\textbf{key\_t} key, \textbf{node\_t} \textbf{***}p\_prev, \textbf{node\_t~**}p\_curr, \textbf{node\_t~**}p\_next, \textbf{bool} srch)} {
  \textbf{node\_t **}prev, \textbf{*}curr, \textbf{*}next\;
  prev\_next = \textbf{nullptr}\;\label{again}
  prev = \&Head\;
  curr = Head\;
\While {\upshape curr != \textbf{nullptr}} {
    next = curr->Next\;
    \If {\upshape !isMarked(next)} {
        \lIf {\upshape curr->Key $\ge$ key} {
            \Break
        }
        prev = \&curr->Next\;
        prev\_next = next\;
    }
    curr = getUnmarked(next)\;
  }
  \If(\tcp*[f]{Search(): skip}){\upshape prev\_next \&\& prev\_next != curr \&\& !srch} {
        \lIf {\upshape !CAS(\&prev, prev\_next, curr} {
            \Goto{again}
        }
        \tcp{Delete prev\_next...curr chain}
    }
  *p\_curr = curr\;
  *p\_prev = prev; *p\_next = next\;
  \Return curr \&\& (curr->Key==key)\;
}
\end{algorithm2e}
\end{minipage}
\caption{Harris' list with optimistic traversals (w/o SMR).}
\label{alg:harris}
\end{figure}

In Figure~\ref{alg:harris}, we present Harris' list without any SMR.
The original Harris' list uses $-\infty$ and $+\infty$ sentinel nodes. We encode the first (pre-head) sentinel implicitly via \texttt{\&Head} (L34); this removes one sentinel node without
changing the semantics: Traversal still starts from \texttt{Head} and CAS acts on \emph{link pointers} (\texttt{node\_t**}). We keep a \emph{pointer} to a link in \texttt{prev} so that CAS can update the predecessor field directly, including \texttt{\&Head}, making the first sentinel unnecessary. Using \texttt{prev = \&curr->next} allows the algorithm to perform \texttt{CAS(prev, curr, new)} uniformly, whether \texttt{prev} refers to \texttt{Head} or to an internal node. We use \texttt{node\_t***} only to return the predecessor's link address (\texttt{node\_t**}) to the caller, since \texttt{CAS} operates on the address of the variable being updated, exactly as in the original Harris algorithm.

The \verb|Insert| method inserts a new node into the list. We slightly diverge from the original version~\cite{HarrisList} by using an optimization~\cite{Herlihy:2012:AMP} which enables read-only search operations. Before inserting a new node, \verb|Insert| checks if the key of the to-be-inserted node exists via \verb|Do_Find|. The new node gets inserted unless its key is already present in the list. \verb|Delete| removes the node from the list. If the key is found in the list, the node is logically deleted at L21 by marking \verb|next|, and then one attempt is made to unlink it from the list at L22.

In \verb|Search|, the underlying \verb|Do_Find| operation skips L43-L44.  In \verb|Insert| and \verb|Delete|, \verb|Do_Find| locates the right position for the key. If a chain of marked nodes is found, the algorithm attempts to clean it up using CAS at L44. \textbf{If we integrate HP without any changes, L37 may crash.}

\section{Design}
\label{sec:design}

\subsection{Bird's-Eye View}
The crux of the problem is HP's inability to properly track physically deleted nodes in Harris' list (or tagged edges in the Natarajan-Mittal tree) while traversing logically deleted nodes. However, if there were some way to confirm that the following logically deleted node has not yet been physically deleted, optimistic traversals would still be feasible.

In Figure~\ref{fig:SCOT_HL_HP}, N2 is still \emph{safe} to access because it is not physically deleted, i.e., N1 still points to N2. Moreover, according to Figure~\ref{alg:harris}, we always make sure that the node to the left is not logically deleted and thus \emph{never unlink nodes in the middle of the chain}, i.e., we can delete just N2, or both of N2 and N3, but we never delete N3 while keeping N2 in the list. In other words, we remove a prefix of the chain.

Combining these observations, we gain a crucial insight that helps solve the issue with HP: It is still safe to access N3 and the following nodes as long as at every step, i.e., after making an HP reservation for the next logically deleted node, we verify that N1 still points to N2. In other words, once we reach N2, we declare it to be a ``dangerous zone.'' Until we reach the end of the chain of consecutive logically deleted nodes, we are going to stay in the dangerous zone and perform additional checks. More specifically, after retrieving N3's pointer and creating its HP reservation, we check that N1 still points to N2. We will continue to perform these checks at every iteration until we reach the end of the chain. If the check fails, we cannot proceed further and need to restart the search operation from the very beginning.

SCOT makes several \textbf{key assumptions}: (1) The data structure supports a \emph{deferred} deletion of ``logically'' deleted nodes and (2) For a logically deleted node to be physically unlinked, its predecessor on the traversal path must not itself be logically deleted. While uncommon in trees or lists, unlinking a node with multiple predecessors requires extra care if some predecessors are valid while others are logically deleted.

\begin{figure}
    \centering
    \includegraphics[width=\columnwidth]{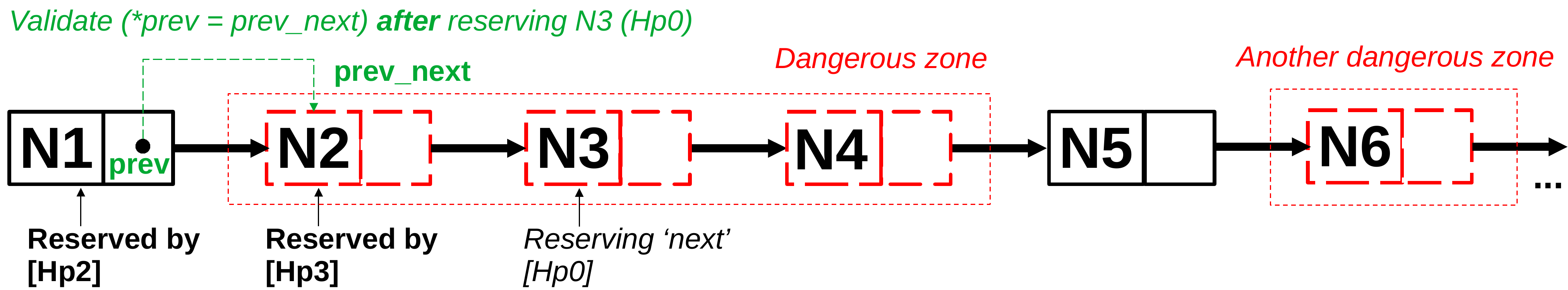}
    \caption{SCOT for Harris' list: validating *prev = prev\_next at every iteration while traversing via the dangerous zone.}
    \label{fig:SCOT_SOLUTION}
\end{figure}

\subsection{SCOT for Harris' List}
\label{sec:scotharis}

In Figure~\ref{fig:SCOT_SOLUTION}, we present our approach for Harris' list. Compared to Michael's approach, we need an additional hazard pointer index (Hp3) which protects the first \emph{unsafe} node, i.e., logically deleted node. This extra hazard pointer prevents the ABA problem, in case the unsafe node gets recycled while we are traversing the list. By holding this extra hazard pointer, we can solely rely on pointer comparison.
In other words, Hp3 protects N2 from our example above irrespective of where we currently are in the chain of logically deleted nodes. Like in Michael's approach, we need Hp2, which protects the \verb|prev| area. We note that in Harris' list, it will be the last \emph{safe}, i.e., not logically deleted node.

We present the pseudocode of SCOT-augmented \verb|Do_Find| in Figure~\ref{alg:scotlist}. We show a simple version (on the left) and a version that unrolls the loop for better performance (on the right).
We use \verb|protect| and \verb|dup|, discussed in Section~\ref{sec:backgroundhp}.
It is crucial to use \verb|dup| such that the old HP index has a lower numerical value than the new HP index (e.g., Hp0 to Hp1). This allows to avoid a small race window when iterating in the same (ascending) order of indices in HP's \verb|retire|. Specifically, when advancing from \verb|next| to \verb|curr|, we copy Hp0 to Hp1, but we cannot do it the opposite way. This ordering is critical because duplicating in the reverse direction (Hp1 $\rightarrow$ Hp0) would create a window  where Hp0 temporarily points to stale data before being safely updated, potentially leading to unsafe access if a concurrent reclamation occurs. By maintaining the ascending order (0$\rightarrow$1, 1$\rightarrow$2, etc.), we ensure that each pointer remains protected throughout the traversal.

Our version on the \emph{left} introduces simple changes to Harris' original list. Specifically, L15 makes sure that the last safe node still points to the first unsafe node while traversing.

The tricky part is avoiding L13-L14 (copying to Hp3) for \emph{unmarked} nodes, which would add a memory barrier absent in the Harris-Michael approach, potentially increasing the algorithm's cost. We resolve this (on the \emph{right}) by unrolling and splitting the loop into two phases: (1) \emph{iterating through the safe zone} and (2) \emph{iterating through the dangerous zone}. Phase 1 only duplicates (shifts) \verb|prev| and \verb|curr| hazard pointers from \verb|curr| and \verb|next|, respectively. This is similar to the Harris-Michael algorithm. Upon leaving Phase 1, \verb|curr| is duplicated (Hp3) in L49. In Phase 2, we no longer duplicate \verb|curr| into \verb|prev| (Hp2), which gives us benefit over the Harris-Michael approach in the dangerous zone.

In the dangerous zone of logically deleted nodes, Figure~\ref{alg:scotlist} runs a check in L55 to ensure that the reservation made in L54 is valid. Once the chain (or its subchain) gets unlinked, L55 fails, and we restart the operation from the very beginning. \textbf{\textcolor{lime}{Green}} highlights changes related to SCOT, and \textbf{\textcolor{pink}{pink}} highlights changes specific to the dangerous zone traversal.

\begin{figure}
\begin{minipage}{.495\columnwidth}
\begin{algorithm2e}[H]
\SetAlgoVlined
\Fn{\upshape \textbf{bool} Do\_Find(\textbf{key\_t} key, \textbf{node\_t} \textbf{***}p\_prev, \textbf{node\_t~**}p\_curr, \textbf{node\_t~**}p\_next, \textbf{bool} srch)} {
  \textbf{node\_t **}prev, \textbf{*}curr, \textbf{*}next\;
  prev\_next = \textbf{nullptr}\;\label{againscotun}
  prev = \&Head\;
  curr = hp.protect(Head, Hp1)\;
\While {\upshape curr != \textbf{nullptr}} {
    next = hp.protect(curr->Next, Hp0)\;
    \If {\upshape !isMarked(next)} {
        \lIf {\upshape curr->Key $\ge$ key} {\Break}
        prev = \&curr->Next\;
        prev\_next = next\;
        \tcp{Copy Hp1[curr] to Hp2}
        hp.dup(Hp1, Hp2)\;
        \HiLiGr \tcp{Copy Hp0[next] to Hp3} 
        \HiLiGr hp.dup(Hp0, Hp3)\;
    }
        \tcp{Dangerous zone: check}
    \tcp{the last safe node (N1) still}
    \tcp{points to 1st unsafe (N2)}
    \HiLi\lElseIf {\upshape prev->load() != \HiLi prev\_next} {\Goto{againscotun}}
    curr = getUnmarked(next)\;
    \tcp{Copy Hp0[next] to Hp1}
    hp.dup(Hp0, Hp1)\;
  }
  \tcp{Skip for Search()}
  \If {\upshape !srch \&\& prev\_next \&\& prev\_next != curr} {
        \lIf {\upshape !CAS(\&prev, prev\_next, curr} {
            \Goto{againscotun}
        }
        Do\_Retire(prev\_next, curr)\;
    }
  *p\_curr = curr\;
  *p\_prev = prev; *p\_next = next\;
  \Return curr \&\& (curr->Key == key)\;
}
\Fn{\upshape \textbf{void} Do\_Retire(\textbf{node\_t *}from, \textbf{node\_t *}to)} {
        \Do(\tcp*[f]{Retire the deleted chain}) {(\upshape from != to)} {
            \textbf{node\_t~*}n = getUnmarked(
            from->Next)\;
            smr\_retire(from)\;
            from = n\;
        }
}
\tcp{Hp0 = 0: Next node (next)}
\tcp{Hp1 = 1: Current node (curr)}
\tcp{Hp2 = 2: Last safe node (prev)}
\HiLiGr \tcp{Hp3 = 3: First unsafe node}
\end{algorithm2e}
\end{minipage}
\begin{minipage}{.495\columnwidth}
\begin{algorithm2e}[H]
\setcounter{AlgoLine}{30}
\Fn{\upshape \textbf{bool} Do\_Find(\textbf{key\_t} key, \textbf{node\_t} \textbf{***}p\_prev, \textbf{node\_t~**}p\_curr, \textbf{node\_t~**}p\_next, \textbf{bool} srch)} {
  \textbf{node\_t **}prev, \textbf{*}curr, \textbf{*}next\;
  prev\_next = \textbf{nullptr}\;\label{againscot}
  prev = \&Head\;
  curr = hp.protect(Head, Hp1)\;
  next = hp.protect(curr->Next, Hp0)\;
\While {\upshape \textbf{true}} {
    \Do {\upshape !isMarked(next)} {
    \lIf {\upshape curr \&\& curr->Key $\ge$ key} {\Goto{harclean}}
     prev\_next = \textbf{nullptr}\;
     prev = \&curr->Next\;
     \tcp{Copy Hp1[curr] to Hp2}
     hp.dup(Hp1, Hp2)\;
     curr = getUnmarked(next)\;
     \lIf {\upshape !curr} {\Goto{hardone}}
     \tcp{Copy Hp0[next] to Hp1}
    hp.dup(Hp0, Hp1)\;
    next = hp.protect(
    curr->Next, Hp0)\;
   }
   \HiLiGr\tcp{Copy Hp1[curr] to Hp3}
   \HiLiGr prev\_next = \HiLiGr hp.dup(Hp1,Hp3)\;
    \tcp{Dangerous zone: check}
    \tcp{the last safe node (N1) still}
    \tcp{points to 1st unsafe (N2)}
   \Do {\upshape isMarked(next)} {
    curr = getUnmarked(next)\;
    \lIf {\upshape !curr} {\Goto{harclean}}
     \tcp{Copy Hp0[next] to Hp1}
     hp.dup(Hp0, Hp1)\;
      next = hp.protect(
     curr->Next, Hp0)\;
    \HiLi \lIf {\upshape prev->load() != \HiLi prev\_next} {\Goto{againscot}}
   }
   }
  \tcp{Skip for Search()}
  \If {\upshape !srch \&\& prev\_next \&\& prev\_next != curr}  { \label{harclean}
        \lIf {\upshape !CAS(\&prev, prev\_next, curr} {\Goto{againscot}}
        Do\_Retire(prev\_next, curr)\;
    }
  *p\_curr = curr\;\label{hardone}
  *p\_prev = prev; *p\_next = next\;
  \Return curr \&\& (curr->Key == key)\;
}
\end{algorithm2e}
\end{minipage}
\caption{SCOT for Harris' list with HP: showing Do\_Find only (\textbf{left} side: unoptimized, \textbf{right:} with the unrolled loop).}
\label{alg:scotlist}
\end{figure}

\subsubsection{Recovery Optimization}
\label{sec:optimization}
For simplicity, we have stated that when validation fails, we restart from the beginning of the list. While this is acceptable, this validation check is not always critical: the last safe node may simply point to another node now (e.g., when a new node is inserted or the chain of logically deleted nodes is already eliminated by a concurrent thread). In such cases, we simply escape from the dangerous zone and continue to the new node. We must, however, ensure that the last safe node is still not logically deleted. When it is deleted, the last safe node (\verb|prev|) is in the dangerous zone itself (though still safe to access due to a prior reservation), so we go back to the beginning of the list. In Section~\ref{sec:eval}, Harris' linked list implements this optimization.

\subsubsection{Why Not Change SMR?\nopunct}
In Figure~\ref{alg:scotlist}, we introduced the SCOT logic directly in Harris' list rather than SMR. While it is feasible to extend the HP API, there are major caveats. For example, Figure~\ref{alg:scotlist}'s optimized version calls 2x \verb|dup| outside the dangerous zone, and 1x \verb|dup| in the dangerous zone (not counting \verb|protect| calls in both cases). If we attempt changing SMR, we will end up with the basic version which needs 3x \verb|dup| everywhere, not counting \verb|protect|. An extra \verb|dup| induces a cost, not present with Michael's approach, which needs 2x \verb|dup|.
Moreover, empirically, we found that the recovery optimization (Section~\ref{sec:optimization}) is beneficial for Harris' list. Conversely, in the Natarajan-Mittal tree, the recovery optimization does not help improve performance for various key ranges, primarily because of the hierarchical structure of the tree, i.e., the tree is likely diverging substantially anyway.

Taking all these factors together, it is better to \textbf{adapt a given data structure} based on its internal characteristics.

\subsection{SCOT for the Natarajan-Mittal Tree}

In the Natarajan-Mittal BST~\cite{NMTree}, all \emph{actual} keys and values are located in the leaf nodes, while the internal nodes contain only \emph{routing} keys used to direct the traversal through the tree.
The Natarajan-Mittal BST incorporates a concept similar to Harris' logical deletion through the use of \emph{tagging} (applied to edges or sibling nodes) and \emph{flagging} (applied to leaf nodes).

The Natarajan-Mittal BST makes one crucial insight: a chain of tagged edges can be eliminated with a single CAS operation. Moreover, for the search operation, the chain of tagged edges can be just skipped over.
This also makes this tree somewhat faster than Ellen et al. tree, depending on the exact workload; both trees were previously evaluated in~\cite{HPPP}, and the Natarajan-Mittal tree is almost always faster.

In essence, the Natarajan-Mittal BST makes an optimization similar to Harris' linked list -- an optimistic traversal of the chain of tagged edges. While optimistic traversals are beneficial in terms of performance, they are also \emph{fundamentally} incompatible with HP and many other robust schemes for the same reason as Harris' list is incompatible with HP.

\begin{figure}
        \includegraphics[width=\columnwidth]{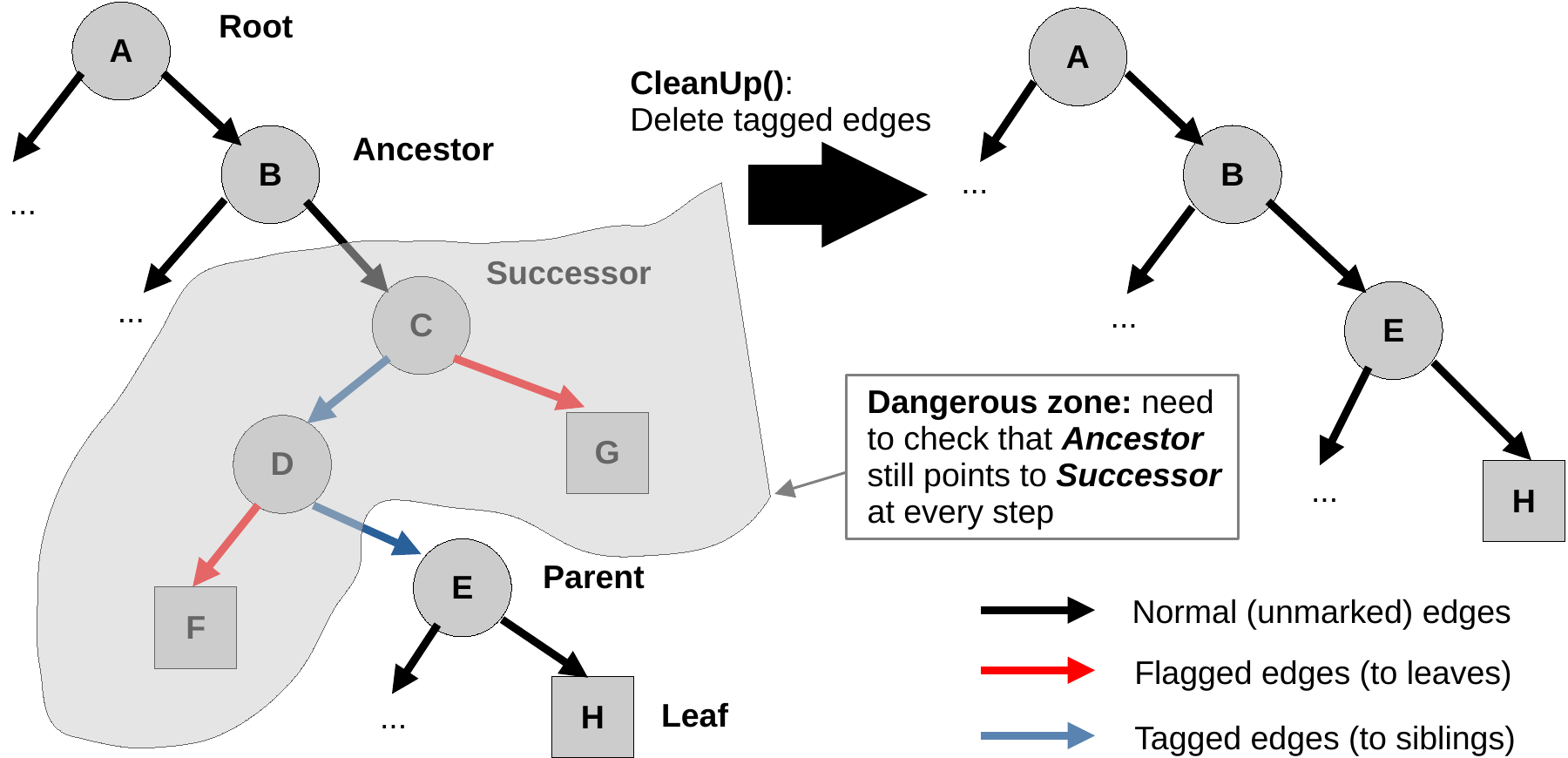}
        \caption{Natarajan-Mittal Tree: Tagged and Flagged Edges.}    
        \label{fig:tree}
\end{figure}

The Natarajan-Mittal BST implements \verb|Insert|, \verb|Delete|, and \verb|Search| operations, which are similar to those of Harris' list, except that the tree traversal is typically faster than a linear search in the list.
\verb|Search| starts at the root and traverses through internal nodes (by comparing keys) and continues until reaching the leaf node with the actual key. If the key is not found in the leaf node, the operation fails.
\verb|Insert| traverses the tree by using a similar, internal \verb|Do_Find| method. If the key is found, \verb|Insert| fails. Otherwise, \verb|Insert| allocates a new internal node, which has \emph{left} and \emph{right} pointers initialized to the leaf node that was found as well as to the leaf node that is about to be inserted. Then the pointer to the old leaf node is being replaced by CAS to the new internal node.
\verb|Delete| also traverses the tree by using \verb|Do_Find|. If the key is not found, it fails. Otherwise,
\verb|Delete| logically deletes the leaf node by changing the corresponding edge's ``flag'' bit. Right after this, the sibling of the flagged leaf is ``tagged''. Finally, an internal \verb|CleanUp| procedure must be called to prune tagged edge(s), as shown in Figure~\ref{fig:tree}. The successor (internal) node will be then replaced with the sibling node.

The SCOT solution is largely similar to that of Harris' list. Per original terminology~\cite{NMTree}, the \emph{successor} node is the node with last untagged edge which comes from \emph{ancestor}. Those can be any distance apart from the \emph{parent} node, which is an immediate parent of the \emph{leaf} node. Tagged edges constitute the ``dangerous zone'' (Figure~\ref{fig:tree}), already discussed in the context of Harris' list; they will reside between the successor and the parent. \emph{At each step, we verify that the ancestor still points to the successor as we traverse via the dangerous zone.}

In the SCOT version of the Natarajan-Mittal BST, five hazard pointers are used, each mapping directly to a specific role in the traversal. 
\textbf{Hp0} protects the current child pointer being followed. 
\textbf{Hp1} protects the current leaf candidate. 
\textbf{Hp2} protects the parent of the leaf. 
\textbf{Hp3} protects the successor node, which is the first node reachable through
an untagged edge and marks the entrance to the tagged (dangerous) zone. 
\textbf{Hp4} protects the corresponding ancestor node whose child pointer should
continue to reference the successor. 

\subsection{Wait-Free Traversals}
\label{sec:waitfreetrav}
The remaining piece is related to EBR's original wait-free traversal guarantees.
With our approach, optimistic traversals may need to occasionally restart the search operation from the beginning if, due to overlapping modifications, the state of the data structure diverges significantly and cannot be recovered locally. This guarantees lock-free but not wait-free progress. This limitation is very similar to that of HP++'s existing solution which also does not support wait-free optimistic traversals due to potential restarts.

However, it is not hard to design an approach which brings wait-free traversals back to HP and other robust schemes without resorting to a full-blown wait-free data structure. (We evaluate Harris' list in Section~\ref{sec:eval} using this approach.)

Our idea is based on a custom fast-path-slow-path method for \verb|Search|. On every iteration, \verb|Insert| and \verb|Delete| check if any searching thread needs help (by running \verb|Help_Threads|). If \verb|Help_Threads| returns true, \verb|Insert| or \verb|Delete| will proceed with the \verb|Slow_Search| slow-path  procedure (see below), which is executed prior to the actual insertion or removal.

Figure~\ref{alg:wf} shows pseudocode. The original \verb|Do_Find| method is still used as is for \verb|Insert| and \verb|Delete| because wait-free guarantees are only provided for traversals. The \verb|Search| operation will run the original \verb|Do_Find| for a few iterations (fast path), after which it will call \verb|Request_Help| to indicate that the thread needs assistance by other threads, via a per-thread record  entry (\verb|thrdrec_t|). \verb|Request_Help| initializes the search key (L28) and then updates \verb|helpTag| (L30), so that other threads are aware that the thread expects help.

On input, \verb|helpTag| contains the current slow-path cycle number; this is needed to prevent any belated updates by helpers intended for past cycles. On output, the same variable, \verb|helpTag|, will contain the result of the search operation. To differentiate inputs and outputs, we reserve one bit (\emph{IsInput}), whereas \emph{Value} represents the actual input tag or output.

The output does not keep track of tags because it is only used by the helpee which is aware of its slow-path cycle. Our approach is specifically tailored for traversals; it avoids non-standard instructions (only a regular CAS) and complexity related to dynamically-allocated descriptors~\cite{10.1007/978-3-642-35476-2_23,10.1145/2555243.2555261}.

Both helpers and the helpee run the same \verb|Slow_Search| operation.
On every iteration, it checks (L34) whether \emph{any} thread has already produced an output (L37), or whether the slow-path input tag has changed since the helper began (L36). \verb|Slow_Search|'s return value is only used by the helpee itself. Since the input tag can only be changed by the \emph{helpee}, the return value in L36 is irrelevant: The L36 condition applies only to \emph{helpers}, who ignore \verb|Slow_Search|'s return value.

\begin{figure}
\begin{minipage}{.51\columnwidth}
\begin{algorithm2e}[H]
\SetAlgoVlined
\textbf{struct} thrdrec\_t \{

    \tcp{=== Private Fields ===}
    \hspace{.5em}\textbf{int} nextCheck\tcp*{= DELAY}
    \hspace{.5em}\textbf{int} nextTid\tcp*{= 0 (Thread ID)}
    \hspace{.5em}\textbf{uint} localTag\tcp*{= 0 (Slowpath \#)}
    \hspace{.5em}\tcp{=== Shared Fields ===}
    \hspace{.5em}\textbf{key\_t} helpKey\tcp*{Key (Input)}
    \hspace{.5em}\tcp{One bit (.IsInput) is reserved to}
    \hspace{.5em}\tcp{differentiate Input/Output}
    \hspace{.5em}\textbf{uint} helpTag\tcp*{= \{.IsInput = 0\}}
\}\tcp*{Tag (Input) or Result (Output)}

thrdrec\_t WF[MAX\_THREADS]\;

\Fn{\upshape \textbf{bool} Help\_Threads(\textbf{key\_t *}p\_key, \textbf{uint *}p\_tag, \textbf{int *}p\_tid)} {

\If {\upshape -{}-WF[TID].nextCheck != 0} {\Return \textbf{False}\;}
WF[TID].nextCheck = DELAY\;
currTid = WF[TID].nextTid\;
WF[TID].nextTid = (currTid + 1) \% MAX\_THREADS\;
\If {\upshape currTid == TID} {\Return \textbf{False}\;}
tag=WF[currTid].helpTag.load();

\lIf {\upshape !tag.IsInput} {\Return \textbf{False}}
key=WF[currTid].helpKey.load();

\lIf {\upshape WF[currTid].helpTag.load() != tag} {\Return \textbf{False}}
*p\_key = key; *p\_tid = currTid\;
*p\_tag = tag\;
\Return \textbf{True}\;
}

\end{algorithm2e}
\end{minipage}
\begin{minipage}{.48\columnwidth}
\begin{algorithm2e}[H]
\setcounter{AlgoLine}{26}
\Fn{\upshape \textbf{uint} Request\_Help(\textbf{key\_t} key)} {
WF[TID].helpKey.store(key)\;
tag = WF[TID].localTag\;
WF[TID].helpTag.store(\{ .Value = tag, .IsInput = 1\})\;
WF[TID].localTag = tag + 1\;
\Return \{ tag, 1 \}\tcp*{Input}
}

\tcp{Note: For the current thread}
\tcp{itself (helpee), helpTid = TID}
\Fn{\upshape \textbf{bool} Slow\_Search(\textbf{key\_t} key, \textbf{uint} tag, \textbf{int} helpTid)} {
\tcp{Like Fig. 5's Do\_Find, but}
\tcp{every iteration (L39, L51)}
\tcp{checks if result is available:}
r=WF[helpTid].helpTag.load();

\If {\upshape r != tag } {
\tcp{Return value is irrelevant}
\tcp{for a different \textbf{input} tag}
\lIf {\upshape r.IsInput} {\Return\textbf{False}}
\tcp{DONE: \textbf{output} value is}
\Return r.Value\tcp*{available}
}

...

...

\tcp{If the key has been found}
ret = \textbf{True}/\textbf{False}\;
\tcp{This thread has found the}
\tcp{result itself => notify others}
WF[helpTid].helpTag.CAS(tag, \{ .Value = ret, .IsInput = 0\})\;
\Return ret\;
}
\end{algorithm2e}
\end{minipage}
\vspace{-12pt}
\caption{Wait-Free Traversals with SCOT.}
\label{alg:wf}
\end{figure}

\section{Correctness}

We analyze the SCOT technique to establish two core properties: (1) \textbf{Safety:} no thread accesses previously reclaimed memory, and (2) \textbf{Lock-Freedom:} the system guarantees global progress.
We also analyze SCOT's wait-free traversal extension (Figure~\ref{alg:wf}). We prove that the $\textsc{Search}(key)$ operation terminates in a finite number of steps for all threads (wait-freedom) while preserving safety and linearizability.

Without loss of generality, for simplicity, we assume a sequentially consistent memory model in our arguments.
Also, Theorems~\ref{thm:traversal-safety}~and~\ref{thm:lockfree} \textbf{consider only Harris' list}, but similar arguments apply to the Natarajan-Mittal tree as well.

\subsection{Memory Bounds}
\begin{theorem}[Bounded Memory Overhead]
\label{theorem:memory}
For a SCOT-based implementation of Harris' list and the Natarajan-Mittal tree, the total memory usage with HP is $O(|D| + N)$, where $|D|$ is the number of live nodes and $N$ is the number of threads.
\end{theorem}

\begin{proof}
Each thread uses a fixed number $H$ of hazard-pointer slots ($H{=}4$ for Harris' list and $H{=}5$ for the Natarajan-Mittal tree), so \emph{at most} $H\cdot N$ nodes are reserved at any time.
Let $R$ be a constant defining a limbo-list scanning threshold (e.g., $128$).
Removed nodes are retired and placed into a per-thread limbo list, and reclamation is triggered only after a thread has retired $R$ nodes. Between reclamation cycles, each thread can accumulate at most $R$ unreclaimed nodes, giving at most $N\cdot R$ additional retired nodes system-wide.

During limbo-list scanning, any retired node not protected by a hazard pointer is freed; at most $H \cdot N$ previously retired nodes remain unreclaimed, and $N\cdot R$ additional nodes can still be in the limbo list before scanning completes. The total number of unreclaimed nodes at any point of time is $HN + NR = O(N)$.
Including $|D|$ reachable nodes, the total memory bound becomes $O(|D| + N)$.
\end{proof}

\subsection{Safe Optimistic Traversal}
\begin{theorem}[HP Compatibility \& Traversal Safety]
\label{thm:traversal-safety}
When a thread $T$ traverses a dangerous zone consisting of the marked sequence $\langle N_k, N_{k+1}, \dots, N_\ell \rangle$, we assign hazard pointers so that $\texttt{HP(T,2)}$ protects the last safe predecessor $N_{k-1}$, $\texttt{HP(T,3)}$ protects the head of the marked region (the first unsafe node $N_k$), and $\texttt{HP(T,1)}$ protects the current node $N_i$ during traversal. Under this assignment, $T$ never dereferences a reclaimed node.
\end{theorem}

\begin{proof}
When $T$ detects the marked region, it protects the last unmarked predecessor $N_{k-1}$ and the first marked node $N_k$, thus making sure that both of them will not be reclaimed.
As $T$ traverses the marked chain, each node $N_i$ and its successor are hazard-protected, and $T$ consistently validates that the predecessor link in $N_{k-1}$ still points to $N_k$.
The risk of unsafe memory accesses exist \emph{only} after physical unlinking.
Any physical unlinking requires the predecessor node to remain unmarked (Figure~\ref{alg:harris}, L22). Thus, only $N_{k-1}$'s link can change, irrespective of the actual removal operation among nodes in the marked sequence.
Due to being reserved, $N_k$ cannot get recycled, and when $N_{k-1}$'s link changes, it points to some other node, causing validation to fail. Traversal restarts (Figure~\ref{alg:scotlist}, L15 and L55) before reaching anything that has already been reclaimed in the marked sequence.
\end{proof}

\subsection{Lock-Freedom}
\begin{theorem}[Lock-Free Updates]
\label{thm:lockfree}
In the SCOT-based design, all update operations (\textsc{Insert} and \textsc{Remove}) are lock-free.
\end{theorem}

\begin{proof}
Each update proceeds through repeated attempts (Figure~\ref{alg:harris}, L11~and~L19). Assuming a finite list length, in each attempt, SCOT runs \textsc{Do\_Find}, which consists of a bounded number of local steps to reach the target location (reads and validation). A restart (Figure~\ref{alg:scotlist}, L15~and~L55)  occurs only when detecting a changed link. The link changes \emph{only} during another successful update: \textsc{Insert} (Figure~\ref{alg:harris}, L15), \textsc{Delete} (Figure~\ref{alg:harris}, L21), or \textsc{Do\_Find} (Figure~\ref{alg:scotlist}, L19~and~L58).

There is one attempt in \textsc{Insert} (L15) or \textsc{Delete} (L21) to change the link after \textsc{Do\_Find}; a failure means that another thread has changed the same location first, indicating global progress. If the current thread indefinitely fails, every failure is caused by a successful update by another thread.
While any thread can starve indefinitely in \textsc{Insert}, \textsc{Delete}, or \textsc{Do\_Find}, there is always at least one thread that makes progress.
\end{proof}

\subsection{Wait-Free Traversals}

\begin{lemma}[Finite Wait Time]
\label{lemma:bfp}
Each thread that called \textsc{Request\_Help} is helped by at least one thread after at most $N \cdot \textsc{DELAY}$ update operations, where $N$ is the number of threads.
\end{lemma}

\begin{proof}
\textsc{Insert} and \textsc{Delete} call \textsc{Help\_Threads} every iteration. They assist exactly \emph{one} thread after \textsc{DELAY} calls of \textsc{Help\_Threads}. (The \texttt{nextCheck} variable in Figure~\ref{alg:wf}, L13 amortizes checks.)
Threads are being served in a circular fashion (L17). Since the number of threads is $N$, the thread under consideration will be served after $N$ calls amortized by the $DELAY$ factor.
\end{proof}

\begin{lemma}[Uniqueness]
\label{lemma:unique}
For any slow-path tag~$t$, only one thread writes an output; no late helper replaces a newer value.
\end{lemma}

\begin{proof} 
A request is posted by \textsc{Request\_Help}, which writes $\mathit{helpTag} \leftarrow \langle v,\mathrm{In}\rangle$ for a per-thread version $v$ and returns $\langle v,\mathrm{In}\rangle$ to the requester. A helper publishes a result only via the atomic transition at L41: 
$\textstyle \mathrm{CAS}\big(\mathit{helpTag},\langle v,\mathrm{In}\rangle,\langle r,\mathrm{Out}\rangle\big)$.
Since CAS compares against the exact input tag $\langle v,\mathrm{In}\rangle$, at most one helper succeeds for the same version~$v$. After that, $\mathit{helpTag}$ changes to $\langle r,\mathrm{Out}\rangle$, so any remaining helper observing $\langle ...,\mathrm{Out}\rangle$ will fail the CAS. Future requests use strictly increasing version numbers ($v' > v$), ensuring that stale helpers fail the CAS on $\mathit{helpTag}$, which is now $\langle v',\mathrm{In}\rangle$.
\end{proof}

\begin{lemma}[Bounded Slow Path]
\label{lemma:bsp}
\textsc{Slow\_Search} terminates after at most $O(N^2\cdot DELAY + \mathit{depth}(L))$ operations, where $\mathit{depth}(L)$ is the maximum depth of the list.
\end{lemma}

\begin{proof}
\textsc{Slow\_Search} may need to restart due to a structural change, which implies that another thread completes an \emph{update} operation.
On each iteration, \textsc{Slow\_Search} re-reads the requester's $\mathit{helpTag}$ (L34) to detect completion by another thread, i.e., when another thread completes CAS in L41, at which point the loop terminates immediately.

Per Lemma~\ref{lemma:bfp}, at least one thread assists the helpee after $N\cdot DELAY$ update operations. This thread may also get stuck in its \textsc{Slow\_Search}. Consequently, another thread after $N\cdot DELAY$ operations will join. After $O(N^2\cdot DELAY)$ operations, \emph{all} threads will be in \textsc{Slow\_Search}, which
will be then unobstructed and conclude after $\mathit{depth}(L))$ steps.
\end{proof}

\begin{theorem}[Wait-Free Search]
\label{thm:waitfree}
\textsc{Search(key)} is wait-free.
\end{theorem}

\begin{proof}
\textsc{Search} executes the fast path for a finite number of restarts $M$, which bounds its execution time to $O(M\cdot depth(L))$. Subsequently, it calls \textsc{Request\_Help} and then switches to \textsc{Slow\_Search}, which is bounded per Lemma~\ref{lemma:bsp}.
\end{proof}

\begin{theorem}[Linearizability]
\label{thm:linear}
\textsc{Search(key)} is linearizable.
\end{theorem}

\begin{proof}
Both the fast and slow paths preserve the same validation and update semantics
as the underlying lock-free data structure. A successful search linearizes at the read that observes a matching key under a valid link predicate, and an unsuccessful search linearizes at the final validated predecessor--successor relation. When help is provided, the helper executes exactly the same steps in the same order.
\end{proof}

\section{Evaluation}
\label{sec:eval}
Our benchmark reuses and substantially extends the test harness from~\cite{HEBenchmark}, which already implements the Harris-Michael list (HMList) and a few SMR schemes: EBR, HP, and HE. We implemented Harris' list (HList) and the Natarajan-Mittal tree (NMTree)
from scratch by carefully considering all known optimizations. We also added IBR~\cite{IBR} and Hyaline-1S~\cite{HYALINEPLDI} SMR schemes into the benchmark. We evaluated the original HP version~\cite{hazardPointers} without extra optimizations. We also implemented and evaluated an optimized HP version (HPopt), 
which captures a local snapshot of shared data prior to the limbo list scanning~\cite{HYALINEPLDI}. HP vs. HPopt shows a substantial difference in some tests. Finally, we implemented a similar optimization for HE and IBR (not applicable to Hyaline-1S), as these schemes also benefit from it.

We did not implement HP++~\cite{HPPP} due to its substantial API differences (i.e., lack of an easy integration). HP++ is already known to perform mostly worse than HP for the \emph{same} data structure~\cite{HPPP}, i.e., HP++ shows some performance benefits only when comparing different data structures (the Harris-Michael HP list vs. Harris' HP++ list). However, this comparison is largely irrelevant as we are now able to implement Harris' list and other data structures in HP directly.

We evaluate HMList, HList, and NMTree under different reclamation strategies: NR (no reclamation, leak memory), EBR, HP, HPopt, IBR, HE, and HLN (Hyaline-1S). For HList and NMTree, we use SCOT when evaluating them for all SMR schemes except NR/EBR. For HList, we present a version with wait-free traversals (Section~\ref{sec:waitfreetrav}); the lock-free version yields almost identical results. These algorithms were previously infeasible for HP/HPopt/IBR/HE/Hyaline-1S~\cite{HPPP, DRC}.

We calibrated all SMR schemes to maximize throughput while minimizing not-yet-reclaimed objects. We found that amortizing limbo list scans at a frequency of one scan per 128 retire calls works well for EBR, HP, HPopt, HE, IBR, and Hyaline-1S. Also, for EBR, IBR, HE, and Hyaline-1S, we set the epoch increment frequency to 12 times the thread count.

We run experiments on Ubuntu 22.04.5 LTS using an AMD EPYC 9754 (Zen~4) system with 128 physical cores (256 hardware threads w/ hyperthreading enabled), 384 GiB of RAM, and a maximum clock speed of 3.10 GHz.
The benchmark is written in C++ and compiled using Clang++ 14.0.0 with -O3 optimizations because its compiled code tends to perform slightly better than the code generated by Ubuntu's (default) GCC 11.4.0. For memory allocation, we use \textbf{Microsoft's \texttt{mimalloc}}~\cite{MIMALLOC} since it scales much better in multi-threaded code compared to glibc's stock malloc. (To stress test algorithms, we also ran all tests with glibc's stock malloc.)

We note that recent AMD EPYC architectures, including the evaluation server, show significantly improved HP performance due to reduced memory barrier overheads, the main prior bottleneck. AMD Zen~3 already introduced key cache-system optimizations that reduced these overheads, and Zen~4 further advances the architecture with improved cache-line sharing and enhanced multithreaded performance~\cite{amdEPYC}. \textbf{Thus, our HP results, especially HPopt, are better than usual and often close to EBR.}

\begin{figure*}
    \begin{subfigure}{0.495\textwidth}
        \includegraphics[width=\textwidth]{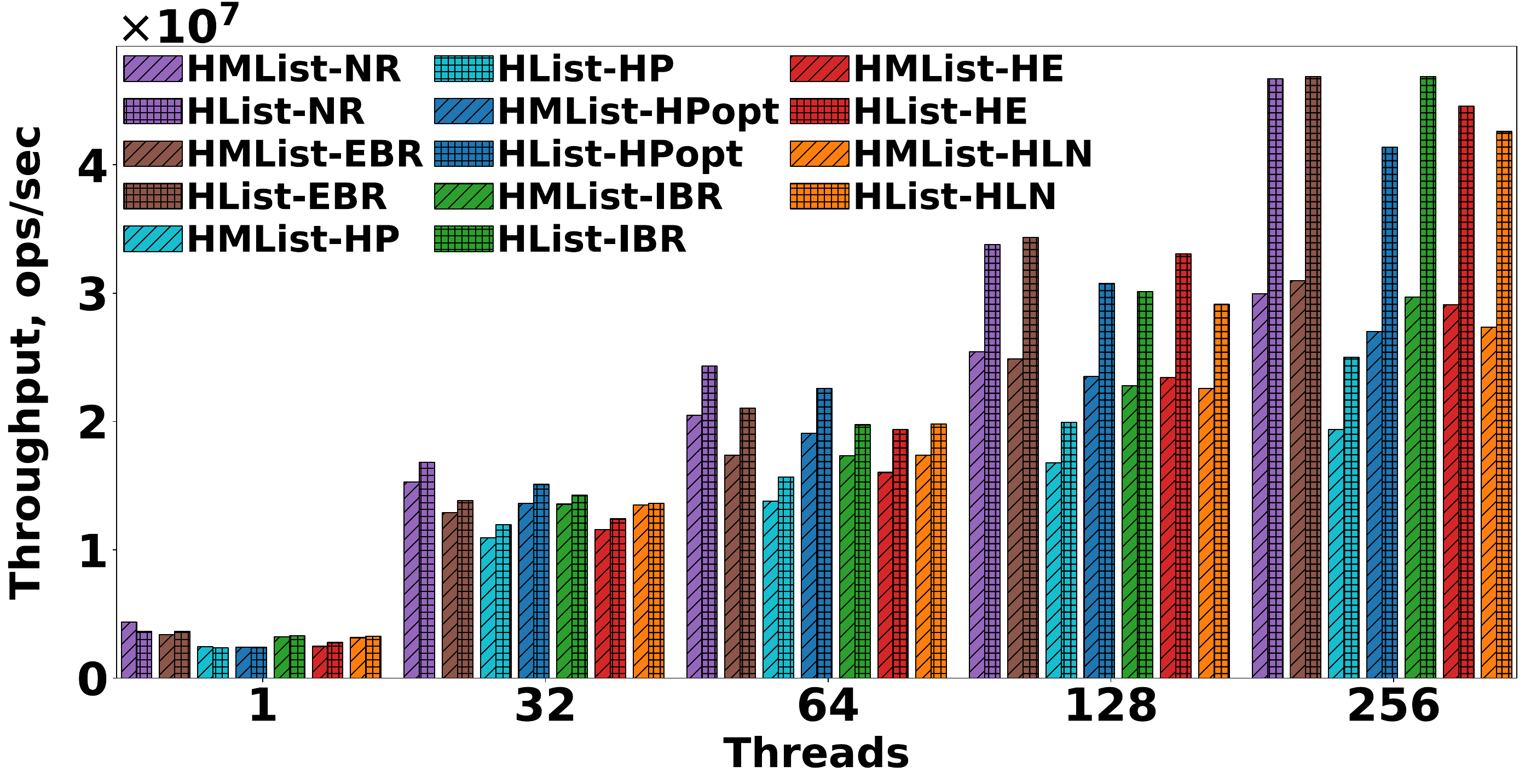}
        \vspace{-15pt}
        \caption{Key Range = 512}
        \label{fig:List_thp_512_RW}
    \end{subfigure}
    \begin{subfigure}{0.495\textwidth}
        \includegraphics[width=\textwidth]{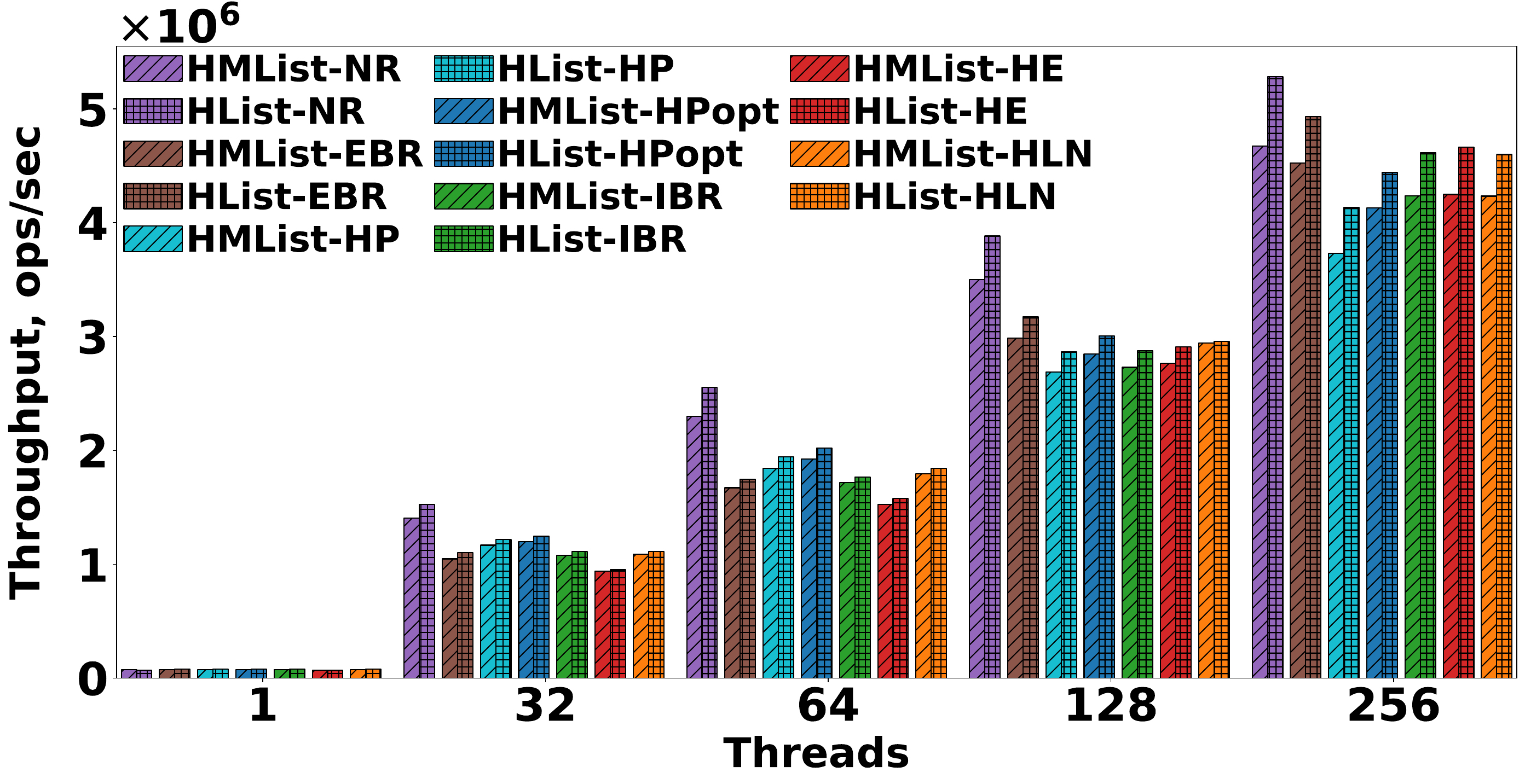}
        \vspace{-15pt}
        \caption{Key Range = 10,000}
        \label{fig:List_thp_10k_RW}
        \end{subfigure}
        \caption{Linked List Throughput (50\% Read - 50\% Write): \emph{higher is better}. (Skipped 384 threads; close to 256.)}
        \label{fig:List_thp_RW}
\end{figure*}

\begin{figure*}
    \begin{subfigure}{0.495\textwidth}
        \includegraphics[width=\textwidth]{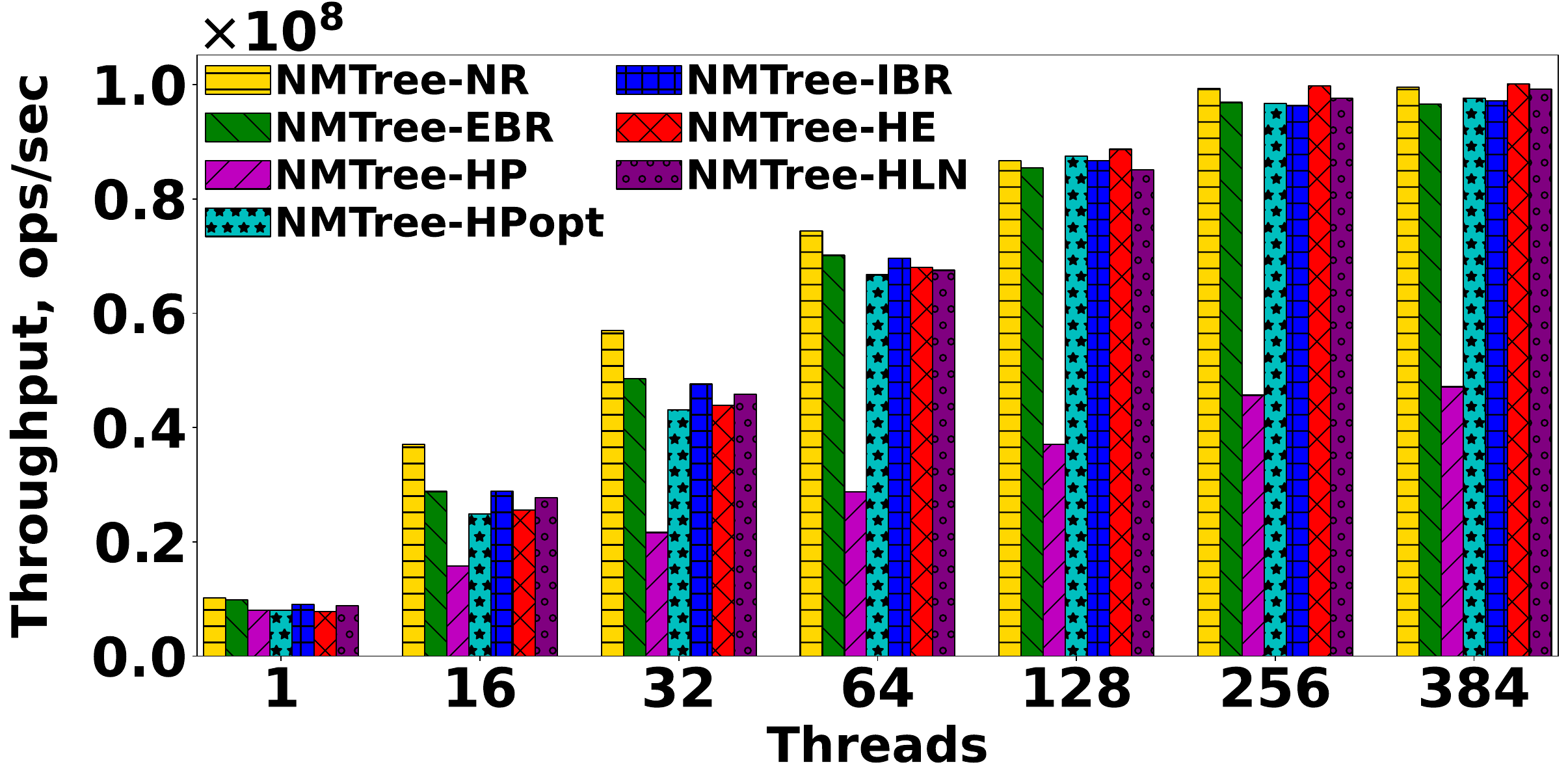}
        \vspace{-15pt}
        \caption{Key Range = 128}    
        \label{fig:Tree_thp_128_RW}
    \end{subfigure}
    \begin{subfigure}{0.495\textwidth}
        \includegraphics[width=\textwidth]{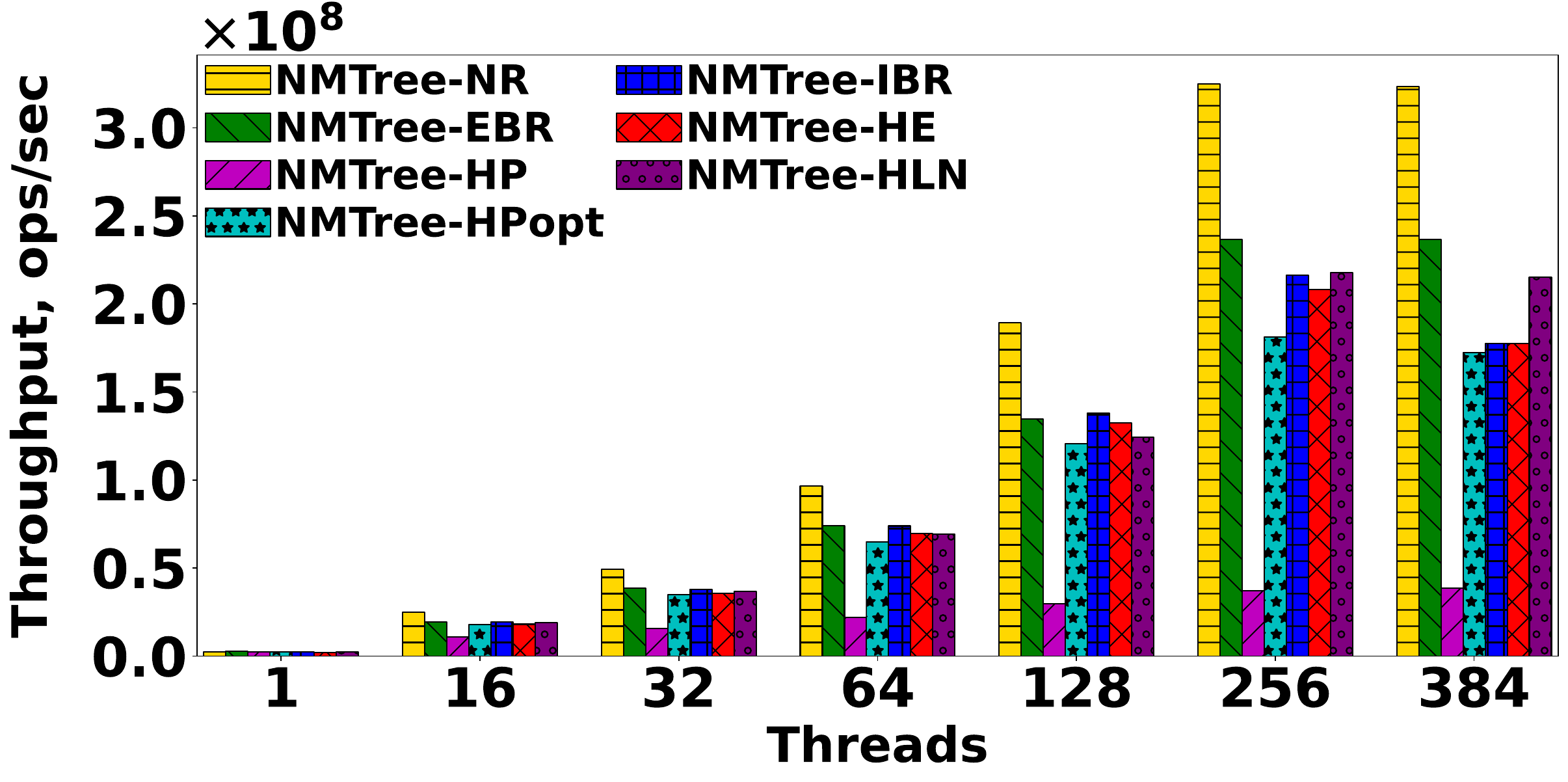}
        \vspace{-15pt}
        \caption{Key Range = 100,000}
        \label{fig:Tree_thp_100k_RW}
    \end{subfigure}
        \caption{NMTree Throughput (50\% Read - 50\% Write): \emph{higher is better}.}
        \label{fig:NM_thp_RW}
\end{figure*}

We measure both throughput and memory overhead. Memory overhead represents the number of not-yet-reclaimed objects, which are nodes that remain in an unreclaimed state due to the use of SMR. To calculate memory waste, we periodically record the average number of not-yet-reclaimed nodes in every thread. \emph{We skip memory overheads for Hyaline-1S due to its global nature of reclamation, making it hard to calculate the number of unreclaimed objects locally per thread using this low-overhead mechanism which works for other SMR schemes.}

Each benchmark begins with prefilling the data structure with unique keys using 50\% of the specified key range. We conduct 5 independent runs, each lasting 10 seconds. The median of all runs is used as the final measure for both throughput and memory overhead. (Across all experiments, the relative standard deviation remains mostly below 1\%.) Our experiments use 1, 16, 32, 64, 128, 256, and 384 threads to evaluate performance for different levels of concurrency (384 threads represent oversubscription). Additionally, we test varying key ranges based on the data structure. For list, we use key ranges of 512 and 10,000. For tree, we use key ranges of 128, 100,000, and 50,000,000. We present results for a common workload, 50\% read - 50\% write (mixed reads and writes). We also measured 90\% read - 10\% write (read-dominated) and 50\% insert - 50\% delete (write-only) workloads, but we omit them as they exhibit largely similar trends.

All throughput figures include a special NR (\emph{no reclamation}) baseline which simply leaks memory and demonstrates the ``upper bound'' for performance. In some cases, EBR and other SMRs can outperform this baseline, e.g.,  when the cost of reclaiming memory is much lower than that of performing fresh allocations. \emph{We also noticed that NR may outperform SMRs, including EBR, under oversubscribed scenarios.}

In Figure~\ref{fig:List_thp_RW}, we present throughput results for the Harris-Michael list and Harris' list under different SMR schemes in a workload consisting of 50\% read and 50\% write operations. We observe that Harris' list consistently outperforms the Harris-Michael list for lower key ranges (512). As the key range increases to 10,000, the performance gap narrows, but Harris' list still maintains its throughput advantage. IBR, Hyaline-1S, HE, and HPopt show excellent performance on Harris' list with the SCOT traversal, often approaching that of EBR. The original HP scheme demonstrates lower throughput than its optimized counterpart (HPopt) in all tests.

\begin{figure*}
    \begin{subfigure}{0.495\textwidth}
        \includegraphics[width=\textwidth]{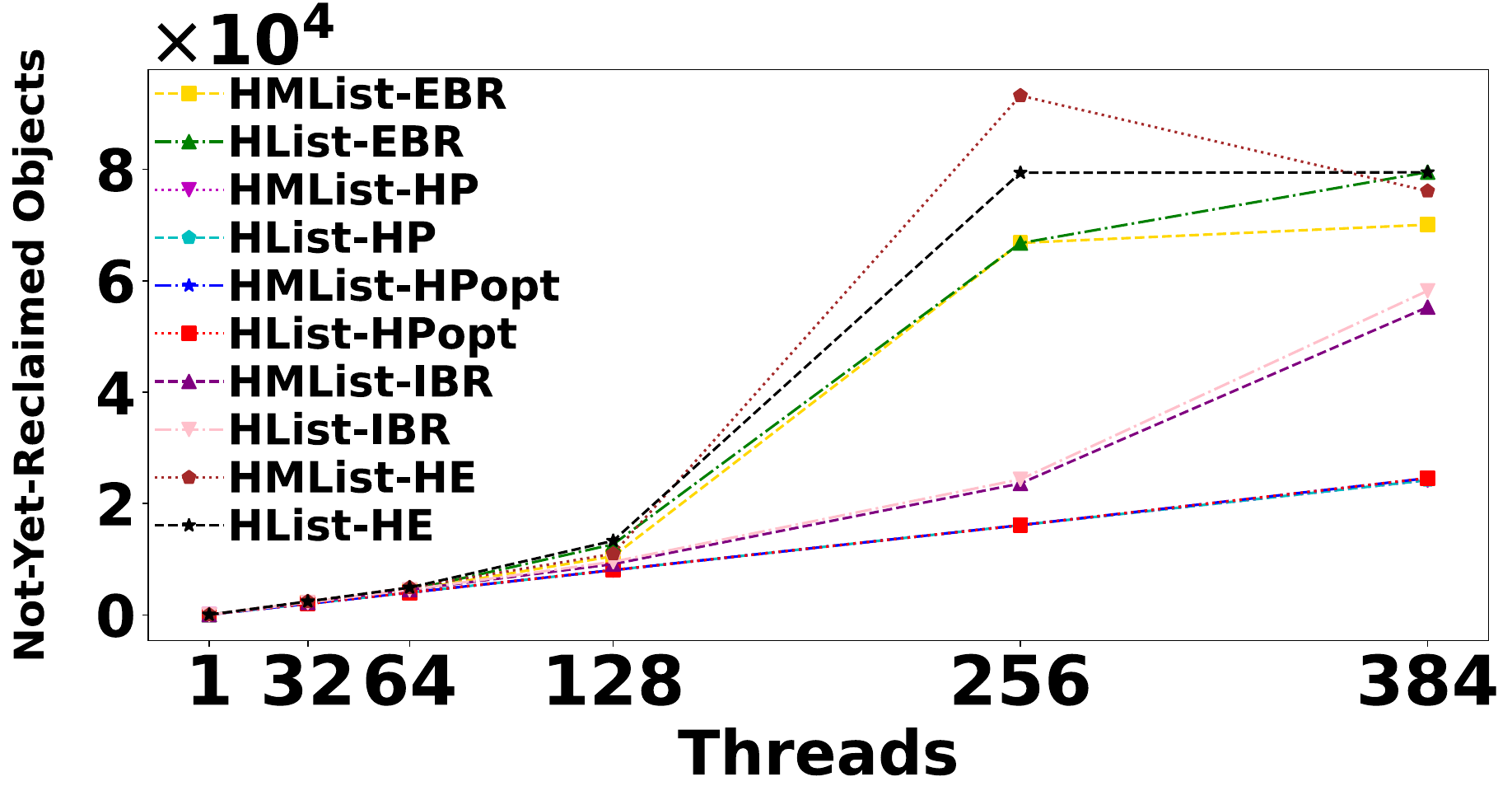}
        \vspace{-15pt}
        \caption{Key Range = 512}
        \label{fig:List_mem_512_RW}
    \end{subfigure}
    \begin{subfigure}{0.495\textwidth}
        \includegraphics[width=\textwidth]{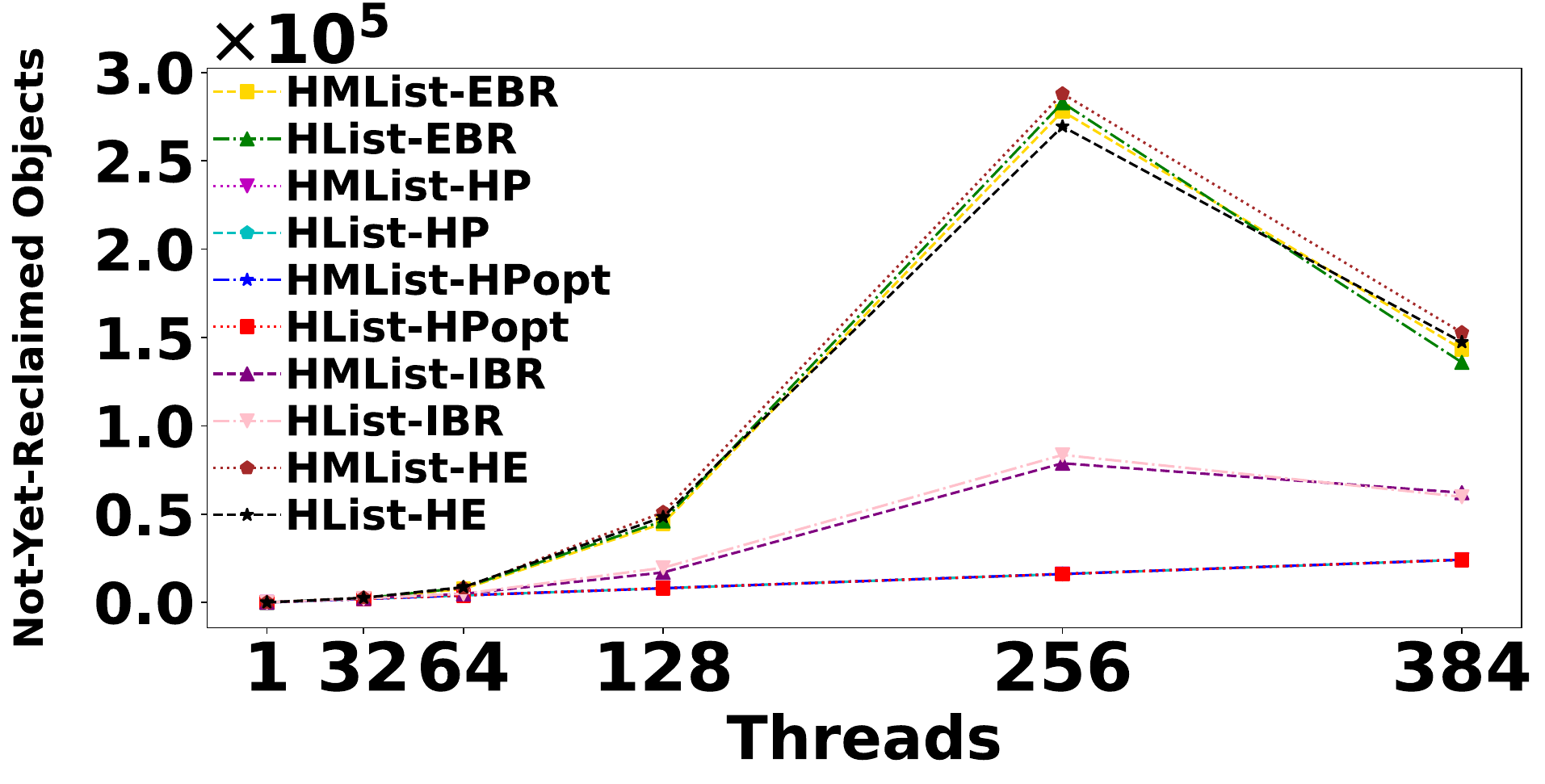}
        \vspace{-15pt}
        \caption{Key Range = 10,000}
        \label{fig:List_mem_10k_RW}
    \end{subfigure}
        \vspace{-5pt}
        \caption{Linked List Average Number of Not-Yet-Reclaimed Objects (50\% Read - 50\% Write): \emph{lower is better}.}
        \label{fig:List_mem_RW}
\end{figure*}

In Figure~\ref{fig:List_mem_RW}, we measure the memory overhead by showing the number of retired but not-yet-reclaimed nodes for Harris-Michael and Harris' lists. As expected, both HP and HPopt consistently maintain the smallest number of such unreclaimed objects compared to all other SMR schemes.

Figure~\ref{fig:NM_thp_RW} presents the throughput of the Natarajan-Mittal tree under a 50\% read and 50\% write workload. For the smaller key range (128), the baseline performance (without reclamation) is comparable to that of the various reclamation schemes, with only a small performance gap observed. In contrast, Figure~\ref{fig:Tree_thp_100k_RW} highlights the scalability of the tree at higher key ranges, where the Natarajan-Mittal tree achieves up to \textbf{240 million operations per second} at 256 threads using the EBR scheme. Hyaline-1S also performs competitively, reaching \textbf{210 million operations per second}, making it the closest in performance to EBR in this configuration.

\begin{figure*}
    \begin{subfigure}{0.495\textwidth}
        \includegraphics[width=\textwidth]{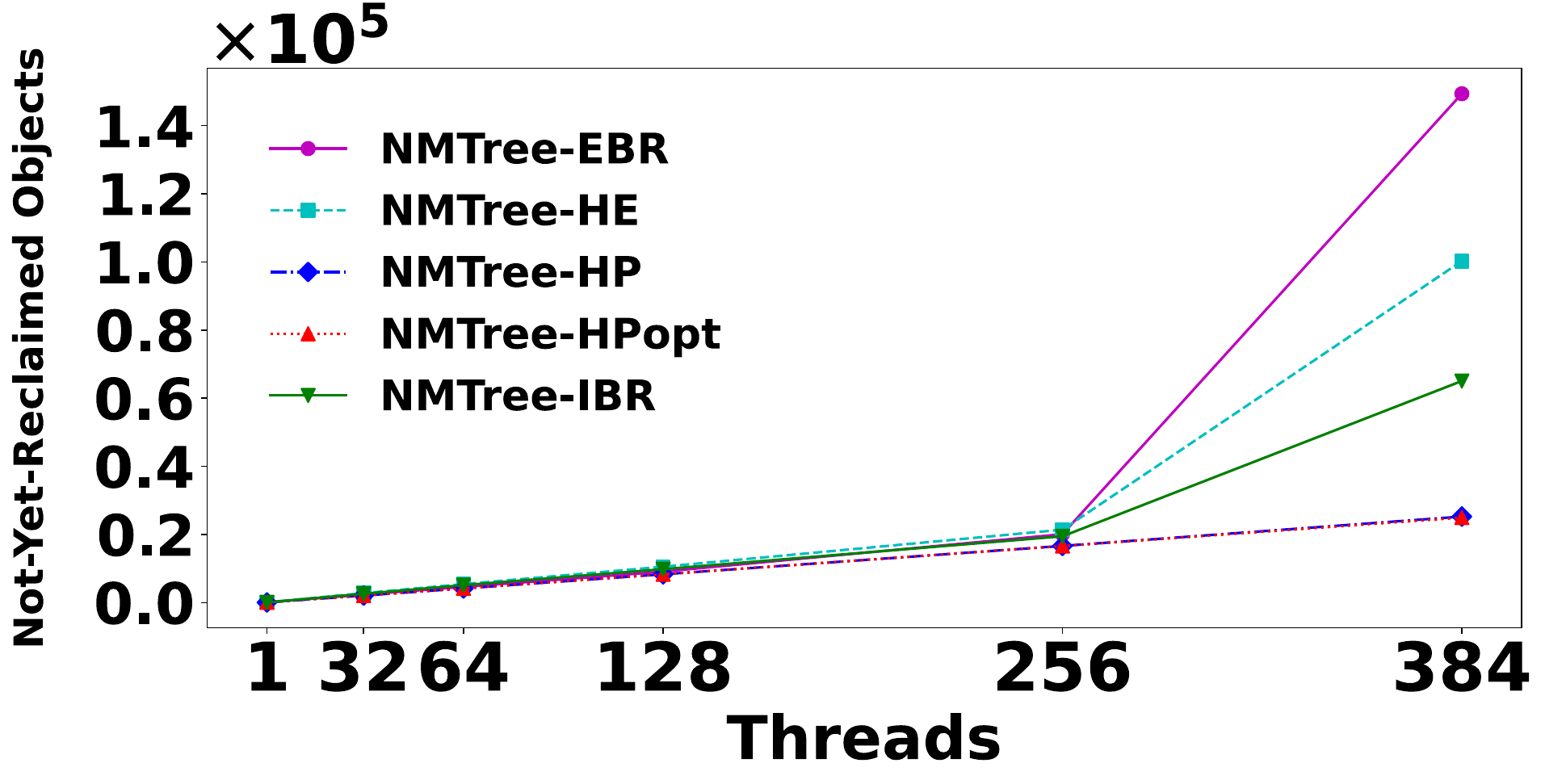}
        \vspace{-15pt}
        \caption{Key Range = 128}    
        \label{fig:Tree_mem_128_RW}
    \end{subfigure}
    \begin{subfigure}{0.495\textwidth}
        \includegraphics[width=\textwidth]{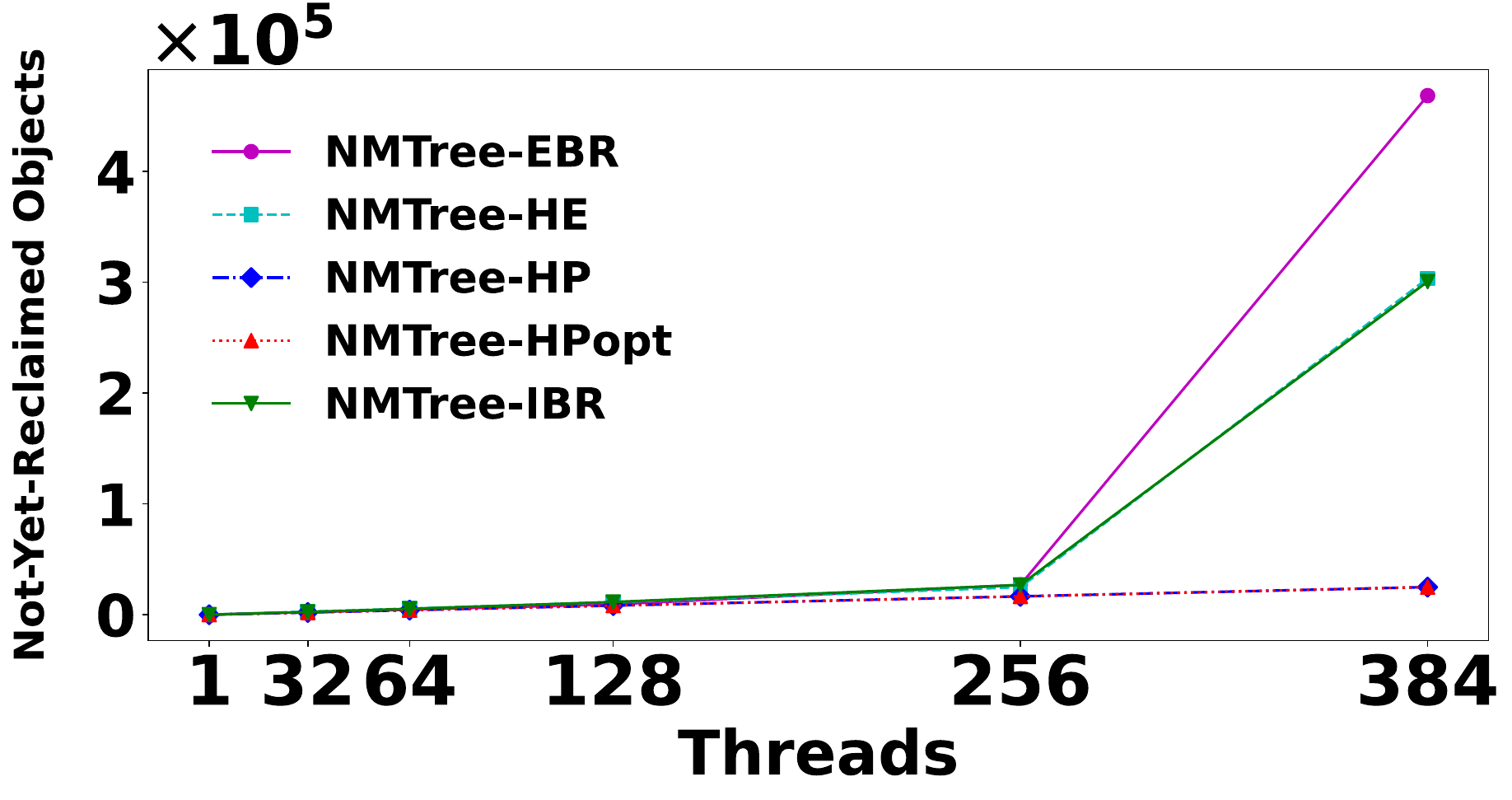}
        \vspace{-15pt}
        \caption{Key Range = 100,000}
        \label{fig:Tree_mem_100k_RW}
    \end{subfigure}
        \vspace{-5pt}
        \caption{NMTree Average Number of Not-Yet-Reclaimed Objects (50\% Read - 50\% Write): \emph{lower is better}.}
        \label{fig:NM_mem_RW}
\end{figure*}

Figure~\ref{fig:NM_mem_RW} presents the memory overhead of the Natarajan-Mittal tree under a 50\% read and 50\% write workload. In both Figure~\ref{fig:Tree_mem_128_RW} and Figure~\ref{fig:Tree_mem_100k_RW}, we observe consistent trends in memory usage. HP and HPopt maintain the lowest number of nodes in the limbo state, which we attribute to its strict and conservative reclamation guarantees. In contrast, EBR exhibits the highest memory overhead, reflecting its more relaxed and delayed reclamation behavior.

\begin{figure*}
    \begin{subfigure}{0.495\textwidth}
        \includegraphics[width=\textwidth]{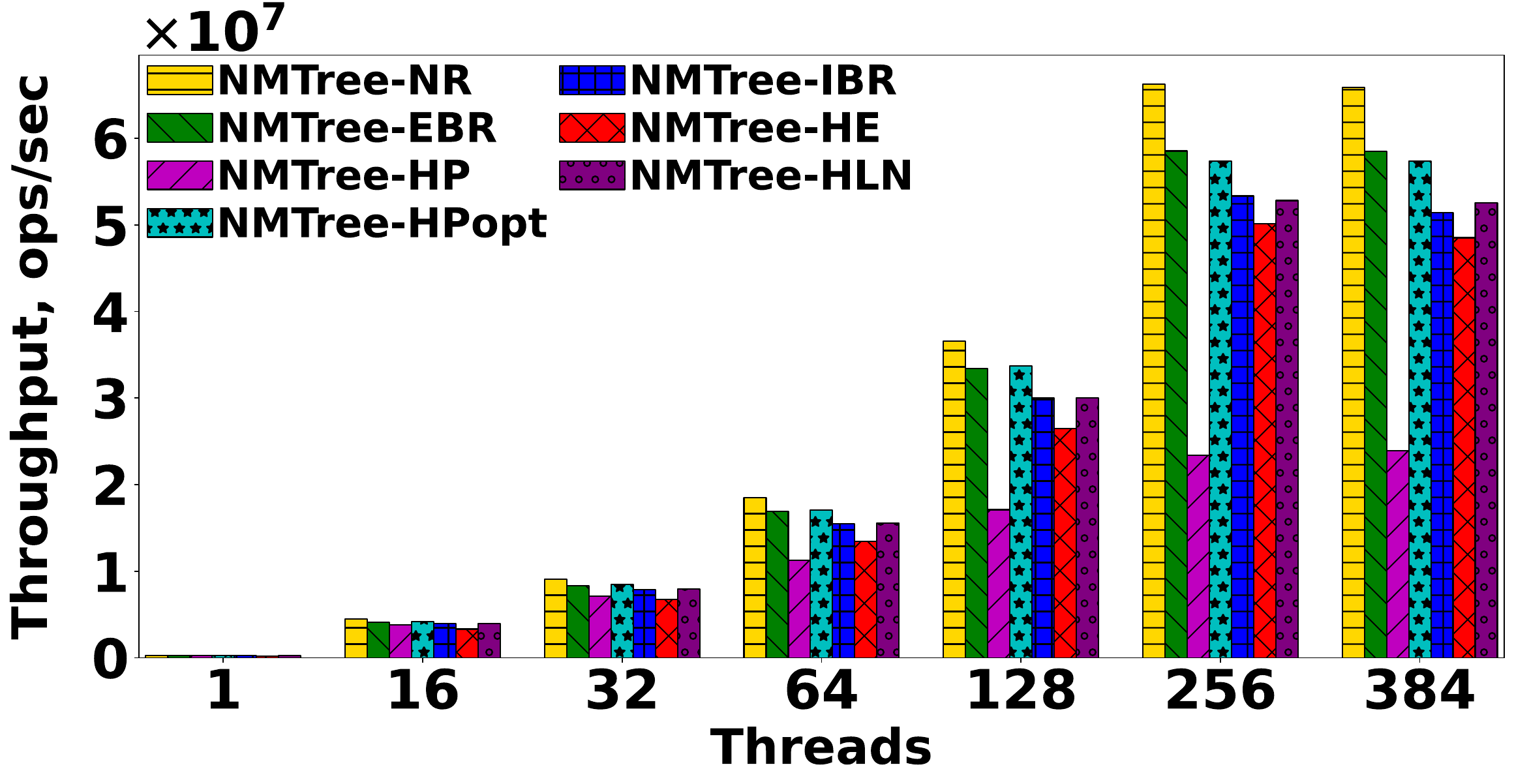}
        \vspace{-15pt}
        \caption{Throughput (\emph{higher is better})}
        \label{fig:Tree_thp_50M_RW}
    \end{subfigure}
    \begin{subfigure}{0.495\textwidth}
        \includegraphics[width=\textwidth]{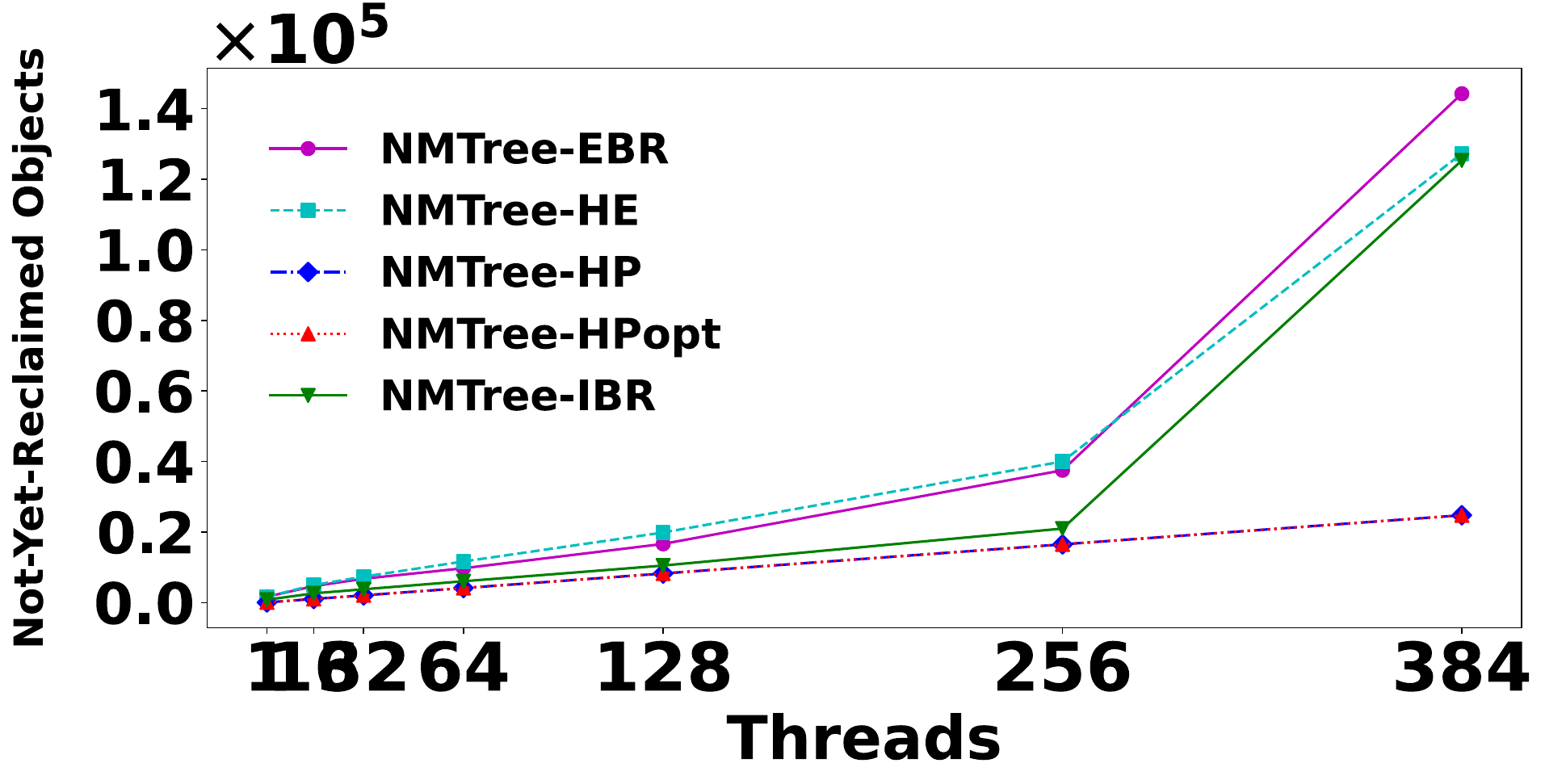}
        \vspace{-15pt}
        \caption{Not-Yet-Reclaimed Objects (\emph{lower is better})}
        \label{fig:Tree_mem_50M_RW}
    \end{subfigure}
    \vspace{-5pt}
    \caption{NMTree (50\% Read - 50\% Write), Key Range = 50,000,000 (not fitting in L1/L2/L3 cache).}
    \label{fig:Tree_thp_mem_50M_RW}
\end{figure*}

We additionally evaluate the Natarajan-Mittal tree on a large key range of 50 million keys, so that the tree does not fit in the CPU cache. Figure~\ref{fig:Tree_thp_mem_50M_RW}a shows that the overall throughput decreases relative to smaller key ranges as expected due to the increase in the traversal depth and reduced cache locality. However, relative trends of SMR schemes remain consistent with the smaller-key experiments. In particular, IBR and Hyaline-1S remain highly competitive, often approaching the throughput of EBR even at high thread counts. 
Figure~\ref{fig:Tree_thp_mem_50M_RW}b reports the average number of not-yet-reclaimed objects. At the highest thread count (384), EBR exhibits the largest number of unreclaimed objects. In contrast, HP and HPopt maintain the lowest memory footprint, reflecting their strict and robust reclamation behavior. 

\begin{table}
    \caption{Restart Statistics for HP, Key Range = 10{,}000.}
\label{tab:restart_stats_HML}
    \vspace{-8pt}
    \renewcommand{\arraystretch}{1.3}
    \centering
    \resizebox{\columnwidth}{!}{
    \begin{tabular}{|c|c|c|c|c|c|c|}
    \hline
    & \multicolumn{3}{c|}{\textbf{The Harris-Michael List}} & \multicolumn{3}{c|}{\textbf{Harris' List}} \\
    \hline
    \textbf{Threads} & \textbf{1} & \textbf{64} & \textbf{256} & \textbf{1} & \textbf{64} & \textbf{256} \\
    \hline
    \textbf{Restarts (per 10 sec)} & 0 & 637938 & 3373950 & 0 & 6 & 59 \\
    \hline
    \textbf{Ops/sec} & 74992 & 1925414 & 4119613 & 78290 & 2025492 & 4424534 \\
    \hline
    & $\mathbf{0\%}$ & $\mathbf{3.31\%}$ & $\mathbf{8.19\%}$ & \multicolumn{3}{c|}{$\mathbf{\approx 0\%}$} \\
    \hline
    \end{tabular}
    }
\end{table}

Finally, in Table~\ref{tab:restart_stats_HML}, we provide empirical restart behavior for the Harris-Michael list and Harris' list with HP for the scenario in Figure~\ref{fig:List_thp_10k_RW}. We observe that for the Harris-Michael list, the restart rate increases from 0\% at 1 thread to 8.19\% at 256 threads. In contrast, Harris' list exhibits $\approx$ 0\% restart rate across any thread count. Since restarts are expensive and require re-traversing the list from the head, they incur substantial overheads. This, combined with a reduced number of CAS during physical unlinking, explains why Harris' list consistently outperforms the Harris-Michael list in Figure~\ref{fig:List_thp_RW}.

\section{Related Work}
\label{sec:related}

\subsection{Non-Blocking SMR}

The ERA theorem~\cite{ERATHEOREM} demonstrates that at most two out of three properties -- (A) \textbf{robustness}, (B) \textbf{easy integration}, and (C) \textbf{wide applicability} -- can be achieved simultaneously. Moreover, \textbf{strong applicability} currently seems to be available only with SMR schemes that implement some kind of quiescence, similar to EBR. However, quiescence periods inevitably lead to potentially unbounded memory usage.

DEBRA+~\cite{debra}, QSense~\cite{Balmau}, ThreadScan~\cite{threadScan}, ForkScan~\cite{forkScan}, NBR~\cite{NBR}, propose a solution to EBR's robustness (A) problem. However, they all need special OS support, such as scheduler interaction or signal delivery, which requires locks in typical OSs. These schemes satisfy (A) and (C) properties but not (B), which is essential to the strict non-blocking progress.
Other schemes that satisfy (A) and (C), e.g., VBR~\cite{VBR}, need non-trivial roll-back mechanisms, which are hard to generalize under a uniform library API.

There were also recent attempts to use automatic garbage collection techniques. Generally speaking, garbage collection is subject to the same trade-offs and limitations, discussed above.
FreeAccess~\cite{FreeAccess} addresses the problem with (C) and optimistic traversals but forgoes (B) by requiring to divide code into
separate read and write phases. OrcGC~\cite{OrcGC} is another lock-free garbage collector. OrcGC requires mechanisms similar to smart pointers either in the compiler or the language itself; it can also be slower than HP in certain tests.

Other SMR schemes such as interval-based reclamation (IBR)~\cite{IBR}, hazard eras (HE)~\cite{HEBenchmark}, wait-free eras (WFE)~\cite{WFE}, Hyaline~\cite{HYALINEPLDI}, and Crystalline~\cite{CRYSTALLINE} provide an HP-like interface; they still lack the wide-applicability property (C) but are often faster than HP. SCOT's method to mitigate the problem with (C) is suitable to these SMR schemes as well.

Drop the Anchor (DTA)~\cite{DTA} is a lightweight memory reclamation technique designed specifically for Harris-style singly linked lists. DTA's idea is to reduce the high per-pointer cost of hazard pointers by amortizing protection over multiple node accesses. Instead of protecting every node, DTA periodically records an anchor every few traversed nodes and relies on timestamps, freezing, and a recovery procedure to ensure safety even in the presence of thread failures. By reducing memory barriers and CAS operations, DTA achieves significantly better performance for traversal-heavy workloads. However, its design is closely tied to the linear traversal pattern of linked lists and depends on the ability to freeze and reconstruct contiguous sublists during recovery, making it difficult to adapt to data structures with branching or non-linear access patterns such as trees, skip lists, or hash maps. In contrast, SCOT enables optimistic traversal with the existing robust SMR schemes by modifying the data structure itself to safely validate traversal paths. Thus, while DTA optimizes the HP mechanism for a specific purpose, SCOT expands the applicability of general-purpose SMRs by adapting high-performance non-blocking data structures to remain compatible, preserving both performance and correctness without requiring SMR-specific recovery logic.

Pass-the-buck~\cite{PTB1, PTB2} uses a similar model as HP. Beware \& Cleanup (B\&C)~\cite{BAC} enhances HP with a limited form of reference counting (RC) for retired records. The key insight is to free a record only when no other records point to it (RC is zero) and no hazard pointer protects it. While potentially addressing optimistic traversals, B\&C's retirement algorithm is complex and incurs higher overhead than HP or RC~\cite{debra}. 

\subsection{Non-Blocking Data Structures}

Timothy L. Harris \cite{HarrisList} was the first to present a \emph{practical} lock-free linked list. Maurice Herlihy and Nir Shavit showed~\cite{Herlihy:2012:AMP} that read-only optimistic search operations are feasible with Harris' list. Unfortunately, Harris' list is incompatible with HP~\cite{OrcGC,FreeAccess,HPPP,DRC,CRYSTALLINE,HYALINEPLDI}. Maged M. Michael proposed~\cite{10.1145/564870.564881} disabling optimistic traversals by unlinking logically deleted nodes one by one, making it feasible to use the algorithm with HP. SCOT enables support for Harris' original list with HP, HE, IBR, and Hyaline, while using optimistic traversals.

A skip list provides probabilistic logarithmic-time search, insertion, and deletion.
Lock-free skip lists~\cite{Herlihy:2012:AMP,epoch1} resemble Harris and Harris-Michael lists but consist of multiple sublists.
However, challenges related to logical deletion and optimistic traversals are identical.

Hash maps store and retrieve key-value pairs in constant time. Lock-free hash map~\cite{10.1145/564870.564881} is an array of  lists and can be based on either Harris' or the Harris-Michael approach.

Binary search trees are also commonly used for efficient search, insertion, and deletion operations. Ellen at al. tree~\cite{ELLENTREE} implements one such approach which is compatible with HP. However, the Natarajan-Mittal tree~\cite{NMTree} is significantly faster (around 43\% according to~\cite{HPPP}) due to its ability to remove multiple concurrently tagged edges in one pass.
Prior to our work, as pointed out in~\cite{DRC,HPPP}, the Natarajan-Mittal tree was incompatible with HP, IBR, and other similar schemes.

\section{Conclusion}
We introduced safe concurrent optimistic traversals (SCOT), a new method which enables support for data structures that are incompatible with HP, IBR, HE, Hyaline-1S, and similar SMRs that lack the wide applicability support. Despite prior beliefs~\cite{HPPP} of incompatibility of the Natarajan-Mittal tree and Harris' list with HP, we have demonstrated that not only they are feasible with HP, IBR, HE, and Hyaline-1S, but they can also retain performance benefits when comparing to equivalent data structures without optimistic traversals. We have also presented a  mechanism which returns wait-free traversals to SCOT-based data structures, further eliminating the gap with EBR-based implementations.

While SCOT does not apply universally, a wide range of data structures can now be safely implemented with the above-mentioned schemes. EBR can still be a good choice for API simplicity, but SCOT enables an alternative with good performance and compatibility when strong robustness is required. We hope our work inspires further research and reevaluation of remaining challenges in SMR, as limited applicability can be addressed in other ways.

\section*{Acknowledgements}
A preliminary version of SCOT previously appeared as a brief announcement at SPAA '25~\cite{10.1145/3694906.3743348}.

We would like to thank the anonymous reviewers and our
shepherd Trevor Brown for their insightful comments and
suggestions, which helped greatly improve this paper.

\section*{Data-Availability Statement}
The benchmark and data supporting this paper are available on Zenodo~\cite{paperArtifact}. The most up-to-date source code is also available at \url{https://github.com/rusnikola/scot}.

\appendix

\section{Artifact Description}

\subsection{Abstract}
Our artifact includes: (1) a Linux VM image deployable with VirtualBox; (2) source code of the benchmark, along with all tested data structures and SMR schemes; (3) scripts for running tests and generating charts. We include detailed instructions for replicating the results in README.txt.

\subsection{Artifact check-list (meta-information)}

\begin{itemize}
\item \textbf{Algorithm:} New algorithm, SCOT.
\item \textbf{Program:} Benchmark with HList, HMList, and NMTree.
\item \textbf{Compilation:} clang++ 14.0.0 with -O3 optimizations.
\item \textbf{Binary:} Linux ELF (x86\_64) executables.
\item \textbf{Run-time environment:} Ubuntu 22.04.5 LTS.
\item \textbf{Execution:} The execution time is passed through a program parameter.
\item \textbf{Output:} Section 5 plots produced as PDF/SVG files.
\item \textbf{Experiments:} Scripts run all presented test cases.
\item \textbf{Workflow frameworks used?:} No.
\item \textbf{Publicly available?:} Yes.
\item \textbf{Code licenses:} 2-BSD and 3-BSD.
\end{itemize}

\subsection{Dependencies}

\subsubsection{Hardware dependencies}
Please ensure that your system (VM) is equipped with a sufficient number of CPUs
and ample memory. Our 128-core machine (256 hardware threads) has 384 GiB of
physical RAM, and we recommend at least 128 GiB, especially if running tests
for the NR baseline.

\subsubsection{Software dependencies}
In our setup with Ubuntu, the following packages were installed: git, build-essential, clang, libstdc++-11-dev, libmimalloc-dev, unzip, zip, python3-pip, libcairo2. We also installed the following python3 packages via pip: numpy, matplotlib, cairosvg, pandas.

\subsection{Compilation}
\begin{verbatim}
cd ./scot/SCOT
make
\end{verbatim}

\subsection{Experiment workflow}

\begin{itemize}
\item Compile the benchmark.
\item Run tests. We provide source.sh which runs all tests presented in the
paper.
\item For individual tests, you can also invoke tests directly. For example, for a lock-free list test with EBR (2 seconds, 4 threads):
\begin{verbatim}
./bench listlf 2 512 1 50 25 25 EBR 4
\end{verbatim}

(You can see all options by running ./bench.)

\item Plot the results. See below.

\end{itemize}

\subsection{Evaluation}

HList and NMTree for HP/HPO/HE/IBR/Hyaline are new
implementations that use the SCOT technique; they constitute the main contribution of the paper (both LF and WF versions for HList). Conversely, HMList, which
is unlike HList was already feasible for all SMR schemes,
and EBR/NR schemes (across all data structures) represent existing baselines.

\textbf{Running all tests:}

For full-blown tests (can take up to 10 hours and require a lot of RAM,
e.g., at least 128 GiB is recommended), execute:

\begin{verbatim}
cd ./scot/Scripts
nohup ./source.sh &
\end{verbatim}

For lightweight tests (which run 2 hours and consume less memory), you can skip the NR baseline:

\begin{verbatim}
cd ./scot/Scripts
nohup ./source_lightweight.sh &
\end{verbatim}

Output results will be located in ./scot/Data  for
the corresponding data structure: listlf (lock-free list),
listwf (list with wait-free traversals), tree (Natarajan-Mittal tree).
Also, ./scot/run\_XXX directory will contain intermediate files. Finally, ./scot/Scripts/run.log is the standard output (log) file.

\textbf{Drawing PDF plots:}

\begin{verbatim}
cd ./scot/Scripts
python3 generate_charts.py
\end{verbatim}

This script scans ./scot/Data/*, processes all list (lock-free and wait-free versions)
and tree benchmark outputs, and produces the corresponding SVG and PDF charts
in the *\_charts subdirectories. By default, this command uses all available
thread counts present in the result files, but the `paper' parameter enables paper-optimized settings.

\bibliography{lockfree}

\end{document}